\pdfoutput=1

\documentclass{article}
\usepackage{microtype}
\usepackage{graphicx}
\usepackage{subfigure}
\usepackage{booktabs}
\usepackage{hyperref}
\usepackage{array}
\usepackage{ragged2e}

\usepackage{amsmath}
\usepackage{listings}
\usepackage{xcolor}
\usepackage{multirow}
\usepackage{comment}
\usepackage{tcolorbox}
 \usepackage{booktabs}
\usepackage{booktabs}
\usepackage{multirow}
\usepackage{xcolor}
\usepackage{pifont}
\usepackage{placeins}
\usepackage{ragged2e}
\usepackage{xspace}

\usepackage{acro}
\acsetup{case-sensitive = false}
\DeclareAcronym{AI}{short=AI, long=Artificial Intelligence}

\DeclareAcronym{CNI}{short=CNI, long=Container Network Interface}
\providecommand{\CNI}{\ac{CNI}\xspace}

\DeclareAcronym{LLM}{short=LLM, long=Large Language Model}
\providecommand{\LLM}{\ac{LLM}\xspace}
\providecommand{\LLMs}{\acp{LLM}\xspace}

\DeclareAcronym{MEC}{short=MEC, long=Multi-access Edge Computing}
\providecommand{\MEC}{\ac{MEC}\xspace}

\DeclareAcronym{ML}{short=ML, long=Machine Learning}

\DeclareAcronym{OVN}{short=OVN, long=Open Virtual Network}

\DeclareAcronym{QoS}{short=QoS, long=Quality of Service}
\providecommand{\QoS}{\ac{QoS}\xspace}

\DeclareAcronym{RAN}{short=RAN, long=Radio Access Network}

\DeclareAcronym{VPC}{short=VPC, long=Virtual Private Cloud}

\DeclareAcronym{EPSRC}{short=EPSRC, long=Engineering and Physical Sciences Research Council}
\providecommand{\EPSRC}{\ac{EPSRC}\xspace}

\DeclareAcronym{JOINER}{short=JOINER, long=Joint Open Infrastructure for Networks Research}
\providecommand{\JOINER}{\ac{JOINER}\xspace}

\DeclareAcronym{RIC}{short=RIC, long=Radio Intelligent Controller}

\providecommand{\RICs}{\acp{RIC}\xspace}

\DeclareAcronym{NTN}{short=NTN, long=Non-Terrestrial Networks in telecommunications}
\providecommand{\NTN}{\ac{NTN}\xspace}

\DeclareAcronym{SLO}{short=SLO, long=Service-Level Objective}
\providecommand{\SLO}{\ac{SLO}\xspace}
\providecommand{\SLOs}{\acp{SLO}\xspace}

\DeclareAcronym{PP}{short=PP, long=Pipeline Parallelism}
\providecommand{\PP}{\ac{PP}\xspace}

\DeclareAcronym{TP}{short=TP, long=Tensor Parallelism}
\providecommand{\TP}{\ac{TP}\xspace}

\DeclareAcronym{DP}{short=DP, long=Data Parallelism}
\providecommand{\DP}{\ac{DP}\xspace}

\DeclareAcronym{RTT}{short=RTT, long=Round-Trip Time}
\providecommand{\RTT}{\ac{RTT}\xspace}
\providecommand{\RTTs}{\acp{RTT}\xspace}

\DeclareAcronym{TTFT}{short=TTFT, long=Time to First Token}
\providecommand{\TTFT}{\ac{TTFT}\xspace}

\DeclareAcronym{TPOT}{short=TPOT, long=Time per Output Token}
\providecommand{\TPOT}{\ac{TPOT}\xspace}

\DeclareAcronym{KV}{short=KV, long=Key-Value}
\providecommand{\KV}{\ac{KV}\xspace}

\DeclareAcronym{PoP}{short=PoP, long=Point of Presence, long-plural-form=Points of Presence}
\providecommand{\PoP}{\ac{PoP}\xspace}
\providecommand{\PoPs}{\acp{PoP}\xspace}

\DeclareAcronym{AWS}{short=AWS, long=Amazon Web Services}
\providecommand{\AWS}{\ac{AWS}\xspace}

\DeclareAcronym{EC2}{short=EC2, long=Amazon Elastic Compute Cloud}

\DeclareAcronym{EKS}{short=EKS, long=Amazon Elastic Kubernetes Service}

\DeclareAcronym{S3}{short=S3, long=Amazon Simple Storage Service}

\DeclareAcronym{GCP}{short=GCP, long=Google Cloud Platform}
\providecommand{\GCP}{\ac{GCP}\xspace}

\DeclareAcronym{FPGA}{short=FPGA, long=Field-Programmable Gate Array}

\DeclareAcronym{RFSoC}{short=RFSoC, long=Radio Frequency System-on-Chip}

\DeclareAcronym{HaaS}{short=HaaS, long=Hypervisor-as-a-Service}

\DeclareAcronym{P4}{short=P4, long=Protocol-Independent Packet-Processing Programming Language}

\DeclareAcronym{PTP}{short=PTP, long=Precision Time Protocol}

\DeclareAcronym{SyncE}{short=SyncE, long=Synchronous Ethernet}

\DeclareAcronym{SDN}{short=SDN, long=Software-Defined Networking}

\DeclareAcronym{WAN}{short=WAN, long=Wide Area Network}
\providecommand{\WAN}{\ac{WAN}\xspace}

\DeclareAcronym{SDWAN}{short=SD-WAN, long=Software-Defined Wide Area Network}
\providecommand{\SDWAN}{\ac{SDWAN}\acuse{WAN}\xspace}

\DeclareAcronym{SAN}{short=SAN, long=Storage Area Network}

\DeclareAcronym{NSXT}{short=NSX-T, long=VMware network virtualisation platform}

\DeclareAcronym{TCA}{short=TCA, long=Telco Cloud Automation}
\providecommand{\TCA}{\ac{TCA}\xspace}

\DeclareAcronym{IDAM}{short=IDAM, long=Identity and Access Management}

\DeclareAcronym{ADC}{short=ADC, long=Application Delivery Controller}

\DeclareAcronym{IT}{short=IT, long=Information Technology}

\DeclareAcronym{SSD}{short=SSD, long=Solid-State Drive}

\DeclareAcronym{HDD}{short=HDD, long=Hard Disk Drive}

\DeclareAcronym{OSI}{short=OSI, long=Open Systems Interconnection}
\providecommand{\OSI}{\ac{OSI}\xspace}

\DeclareAcronym{MAAS}{short=MAAS, long=Metal as a Service}
\providecommand{\MAAS}{\ac{MAAS}\xspace}

\DeclareAcronym{NREN}{short=NREN, long=National Research and Education Network}
\providecommand{\NREN}{\ac{NREN}\xspace}

\DeclareAcronym{PSBA}{short=PSBA, long=Public Sector Broadband Aggregation}
\providecommand{\PSBA}{\ac{PSBA}\xspace}

\DeclareAcronym{NDFF}{short=NDFF, long=National Dark Fibre Facility}
\providecommand{\NDFF}{\ac{NDFF}\xspace}

\DeclareAcronym{TCD}{short=TCD, long=Trinity College Dublin}
\providecommand{\TCD}{\ac{TCD}\xspace}

\DeclareAcronym{QUB}{short=QUB, long=Queen's University Belfast}
\providecommand{\QUB}{\ac{QUB}\xspace}

\DeclareAcronym{UCL}{short=UCL, long=University College London}
\providecommand{\UCL}{\ac{UCL}\xspace}

\DeclareAcronym{DC}{short=DC, long=Digital Catapult}
\providecommand{\DC}{\ac{DC}\xspace}

\DeclareAcronym{NPL}{short=NPL, long=National Physics Laboratory}

\DeclareAcronym{LEO}{short=LEO, long=Low Earth Orbit}
\providecommand{\LEO}{\ac{LEO}\xspace}

\DeclareAcronym{GEO}{short=GEO, long=Geostationary Earth Orbit}
\providecommand{\GEO}{\ac{GEO}\xspace}

\DeclareAcronym{MEO}{short=MEO, long=Medium Earth Orbit}

\DeclareAcronym{VLAN}{short=VLAN, long=Virtual Local Area Network}
\providecommand{\VLAN}{\ac{VLAN}\xspace}
\providecommand{\VLANs}{\acp{VLAN}\xspace}

\DeclareAcronym{VPN}{short=VPN, long=Virtual Private Network}

\DeclareAcronym{L2VPN}{short=L2-VPN, long=Layer-2 Virtual Private Network}
\providecommand{\LtwoVPN}{\ac{L2VPN}\acuse{VPN}\xspace}

\DeclareAcronym{NIC}{short=NIC, long=Network Interface Card}
\providecommand{\NIC}{\ac{NIC}\xspace}

\DeclareAcronym{MTU}{short=MTU, long=Maximum Transmission Unit}
\providecommand{\MTU}{\ac{MTU}\xspace}

\DeclareAcronym{RKE2}{short=RKE2, long=Rancher's lightweight Kubernetes distribution}
\providecommand{\RKE}{\ac{RKE2}\xspace}

\DeclareAcronym{NCCL}{short=NCCL, long=NVIDIA Collective Communications Library}

\DeclareAcronym{VM}{short=VM, long=Virtual Machine}

\providecommand{\VMs}{\acp{VM}\xspace}

\providecommand{\cyrenaica}{\texttt{VM1@Southampton}\xspace}

\providecommand{\augusta}{\texttt{VM2@UCL}\xspace}

\providecommand{\equestris}{\texttt{VM2@QUB}\xspace}
\providecommand{\claudiaxi}{\texttt{VM1@TCD}\xspace}

\definecolor{provencolor}{RGB}{100,100,200}   
\definecolor{pendingcolor}{RGB}{196,85,20}    
\definecolor{green}{RGB}{0,100,0}             

\newcommand{\cmark}{\textcolor{provencolor}{\ding{51}}}
\newcommand{\xmark}{\textcolor{pendingcolor}{\ding{55}}}

\usepackage[accepted]{mlsys2025}

\usepackage{iftex}
\ifnum\ifLuaTeX 1\else\ifXeTeX 1\else 0\fi\fi=1
  \usepackage{fontspec}
\fi

\mlsystitlerunning{InferSlice: SLO-driven LLM Inference over a Telco WAN}

\begin{document}

\twocolumn[
    \mlsystitle{InferSlice: A Reference Architecture for SLO-driven \\ LLM Inference over a Telco Wide-Area Network}
    
    
    \mlsyssetsymbol{equal}{*}
    
    \begin{mlsysauthorlist}
    \mlsysauthor{Juan Parra-Ullauri}{inst1}
    \mlsysauthor{Anthony Chiu}{inst1}
    \mlsysauthor{Hari Madhukumar}{inst2}
    \mlsysauthor{Zhaozhou Wu}{inst2}
    \mlsysauthor{Anderson Bravalheri}{inst2}
     \mlsysauthor{Alex Healing}{inst1}
    \mlsysauthor{Andy Corston-Petrie}{inst1}
    \mlsysauthor{Dimitra Simeonideou}{inst2}
    \mlsysauthor{Nektaria Efthymiou}{inst1}
    
    \end{mlsysauthorlist}
    
    \mlsysaffiliation{inst1}{BT Group}
    \mlsysaffiliation{inst2}{University of Bristol}
    
    \mlsyscorrespondingauthor{Juan Parra-Ullauri}{juan.m.parraullauri@bt.com}
    
    \mlsyskeywords{LLM Inference, Network Slicing, Distributed Systems, Telco AI, GPU Scheduling, SLO-driven Orchestration}
    
    \vskip 0.3in
    
    \begin{abstract}

    LLM inference is increasingly deployed as a core service for enterprise and interactive applications, but sovereignty, compliance, latency and resource-locality requirements can limit the suitability of centralised cloud deployments. Telco operators offer geographically distributed compute and network infrastructure for addressing these constraints. However, existing LLM serving frameworks are designed for tightly coupled data centres and provide limited support for coordinating geographically distributed and heterogeneous resources. We present \textbf{InferSlice}, a reference architecture for Service-Level Objective (SLO)-driven LLM inference over a telco infrastructure, integrating network slicing, GPU partitioning, elastic serving and observability. Led by BT in collaboration with the University of Bristol and NVIDIA, and realised on the JOINER UK national-scale wide-area network (WAN), InferSlice is evaluated on service differentiation across tiers, elastic scale-out, client-edge bandwidth as a bottleneck, and the reach of tensor, pipeline, data parallelism, and prefill/decode disaggregation over the WAN. Our results show that, on a general-purpose telco network, tiers yield distinct latency-throughput regimes, and weight streaming shortens scale-out time. Replication is placement-insensitive but memory-bound; pipeline parallelism tolerates near- and mid-cross-site hops, while tensor parallelism and disaggregation fail beyond a single subnet hop. We share practical insights from operating SLO-driven inference on a telco network, lessons learned and open questions, and hope to encourage further research on distributed and heterogeneous LLM inference systems.
    \end{abstract}
]

\printAffiliationsAndNotice{}

\section{Introduction}%
\label{sec:intro}

\LLMs have become a core service for applications ranging from conversational agents to enterprise decision systems and analytics \cite{huang2025sloaware}. As their deployment becomes increasingly widespread, providing scalable, reliable, and low-latency inference has become a first-order systems challenge. \LLM inference is predominantly served from centralised cloud infrastructure~\cite{yu2022orca,zhong2024distserve}, which can introduce structural limitations. First, \textbf{sovereignty and compliance~\footnote{\href{https://cohere.com/blog/state-of-sovereign-ai-adoption-2026}{The State of Sovereign AI Adoption in 2026}, IDC InfoBrief.}}: routing sensitive data through centralised infrastructure can raise data residency, regulatory, and operational control concerns. Second, \textbf{latency}: users or applications distant from available compute regions incur additional network delay before inference computation begins, degrading interactive service quality. Third, \textbf{resource locality}: centralised serving cannot always exploit compute capacity already available closer to demand, including within telecom infrastructure, even when doing so could alleviate both of the constraints above.\acuse{AI}

\begin{figure}[H]
    \centering
    \includegraphics[width=\linewidth]{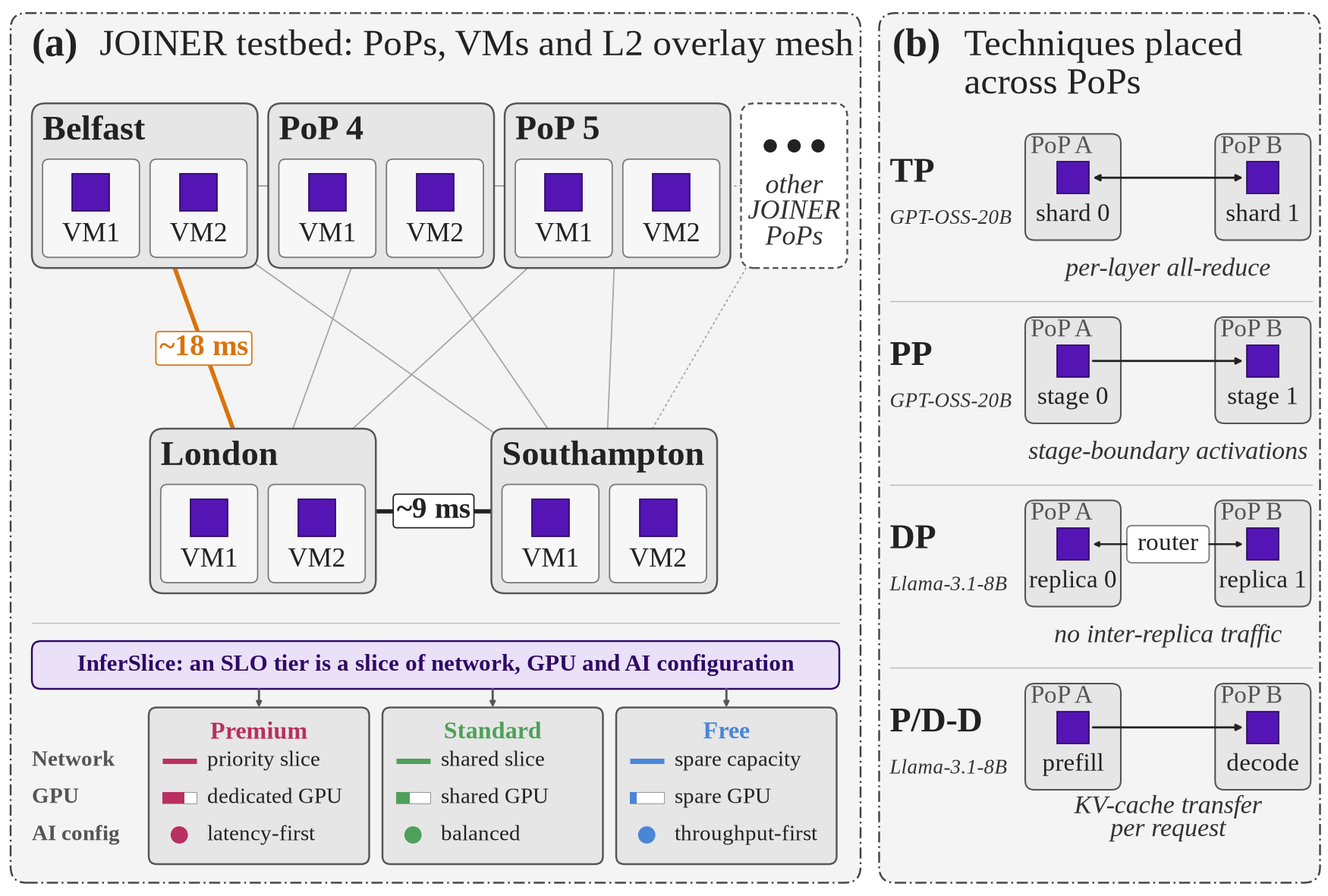}
    \vspace{-0.5cm}
    \caption{(a)~InferSlice deployed on JOINER PoPs with varied network and compute resources. SLO tiers are slices of network, GPU and AI configuration. (b)~Tensor, Pipeline, Data Parallelism (TP, PP, DP) and Prefill/Decode Disaggregation (P/D-D) placed across PoPs.}
    \label{fig:network1}
\end{figure} 
Telco operators offer a compelling substrate for addressing these limitations. Their geographically distributed compute and network infrastructure, already deployed and operated under national regulatory and security frameworks, provides a foundation for serving AI workloads closer to where users and data originate~\citep{taleb2017mec}. Distributed inference across operator-controlled sites can provide shorter network paths, jurisdictional control, and greater operational control and predictability over the network and compute resources on which services depend. This opportunity extends the role of telecommunications infrastructure beyond connectivity toward a distributed platform for AI services.

However, existing \LLM serving frameworks are predominantly designed for tightly coupled data-centre environments~\citep{yu2022orca,zhong2024distserve}, where compute resources are connected by high-bandwidth, low-latency interconnects and can typically be managed as a closely coupled resource pool. Geographically distributed telco infrastructure introduces fundamentally different constraints, including heterogeneous accelerators, variable network latency and bandwidth, and resource availability that varies across sites. These constraints become particularly challenging when serving heterogeneous workloads with different service requirements. Interactive applications such as conversational assistants are sensitive to \TTFT and \TPOT, whereas batch workloads may prioritise sustained throughput and cost efficiency; other applications may require high availability or predictable tail latency, highlighting users and applications requiring different \SLO. Consequently, effective distributed serving requires more than replicating existing inference services across sites: inference configuration, resource allocation, and serving location must be selected according to application \SLOs and the available compute and network resources.

To address this gap, we present \textbf{InferSlice}, a reference architecture for \SLO-driven distributed \LLM inference over geographically distributed telco infrastructure. Led by British Telecom (BT) in collaboration with the University of Bristol and NVIDIA, InferSlice treats multiple telco compute sites as a shared inference substrate and provides the architectural mechanisms required to jointly manage application requirements, distributed compute resources, and network conditions. Rather than assuming a tightly coupled resource pool or a predetermined placement of inference functions, the architecture dynamically matches requests to serving sites and inference configurations according to each application's \SLO and the availability of compute and network resources. InferSlice is realised as a cloud-native platform that integrates distributed \LLM serving, resource orchestration, network management, and end-to-end observability across multiple sites, extending service differentiation from network connectivity to AI inference itself.

We deploy InferSlice on the JOINER\acuse{JOINER} platform~\footnote{\href{https://joiner.org.uk/}{JOINER (Joint Open Infrastructure for Networks Research)}, an international-scale open experimentation platform for future networks research and innovation.}, a UK-\WAN telco testbed spanning multiple cities in the UK, and evaluate its behaviour across both compute and network dimensions. Our experiments examine whether resource and scaling policies can enforce differentiated \SLO outcomes across service tiers under increasing load, whether inference scaling mechanisms can reduce scale-out latency, and how wide-area network latency and bandwidth constrain distributed model parallelism and disaggregation across sites (Figure~\ref{fig:network1}). All experiments are conducted over a shared WAN, without dedicated AI interconnects or RDMA between sites, reflecting real-world conditions in general-purpose telco infrastructure rather than specialised AI-cluster fabrics. Together, these evaluations characterise the feasibility and operating boundaries of \SLO-driven \LLM inference over geographically distributed telco infrastructure and inform the design principles presented in this work.

\section{Background and Related Work}%
\label{sec:background}
\subsection{Distributed LLM Inference Fundamentals}%
\label{sec:background:fundamentals}

Serving LLMs at scale relies on distribution strategies that can be broadly categorised into three largely composable dimensions: \emph{model parallelism}, \emph{replication}, and \emph{functional disaggregation}. While these dimensions are not fully independent, disaggregation in particular alters the internal structure of what would otherwise constitute a single replica. Each addresses a distinct scaling question: model parallelism scales a single model instance across accelerators; replication scales the number of independent model instances; and functional disaggregation scales distinct stages of the inference lifecycle independently of one another.

\subsubsection{Model parallelism.}
When a model does not fit, or does not perform adequately, on a single accelerator, its computation may be partitioned across devices. \TP partitions individual layers' weight matrices across accelerators, requiring frequent collective communication, such as all-reduce, all-gather, or reduce-scatter, between participating accelerators. \PP instead partitions the model by layer across devices, exchanging intermediate activations between pipeline stages and trading reduced synchronisation frequency for pipeline bubble overhead. Expert parallelism (EP) shards mixture-of-experts layers across devices, routing tokens to the accelerators holding the relevant experts. Each of these strategies imposes synchronisation between participating devices at a frequency that typically confines them to low-latency, high-bandwidth interconnect domains such as NVLink or InfiniBand; the binding constraint is synchronisation frequency rather than node or site boundaries per se.

\subsubsection{Replication.}
Often referred to as \DP in inference serving, replication deploys complete, independent copies of a model and distributes incoming requests across them, rather than partitioning a single model instance. Because replicas require no inference-time synchronisation with one another, replication tolerates substantially higher latency and lower-bandwidth interconnects than \TP, \PP, or EP, and is in principle viable across racks, sites, or wide-area network links. Replication therefore provides a natural mechanism for scaling capacity and elasticity: replicas can be added or removed independently as demand changes, while requests can be routed according to load, locality, or \SLO requirements. This independence also provides fault isolation, as congestion or failure affecting one replica need not directly stall others.  

\subsubsection{Functional disaggregation.}
A further scaling dimension separates distinct functions of the inference service into independently provisioned components. The most common example is prefill-decode disaggregation, where the compute-bound prefill phase, responsible for processing the input prompt, is separated from the memory-bandwidth-bound decode phase that generates output tokens autoregressively. These services are connected through a \KV cache transfer mechanism~\citep{zhong2024distserve}. A related form of functional disaggregation externalises \KV-cache management itself, allowing \KV states to be offloaded, persisted, shared, or reused across inference workers. Such decoupling enables resources for different functions to be scaled according to their distinct computational characteristics, at the expense of additional coordination overhead and network dependencies whose feasibility depends on \KV-cache size and available interconnect bandwidth.

\begin{table*}
\centering
\caption{Communication characteristics of distributed LLM inference mechanisms. Communication cost is either \emph{serving-dependent}, fixed once the model and parallelism configuration are chosen, or \emph{workload-dependent}, varying per request with prompt and output characteristics.}
\label{tab:distribution-mechanisms}
\small
\setlength{\tabcolsep}{4pt}
\renewcommand{\arraystretch}{1.15}
\newcolumntype{L}[1]{>{\RaggedRight\arraybackslash}p{#1}}
\begin{tabular}{L{2.8cm}L{2.0cm}L{4.6cm}L{5.5cm}}
\toprule
\textbf{Scaling mechanism} & \textbf{Partition} & \textbf{Communication requirement} & \textbf{Serving vs.\ workload dependency} \\
\midrule
Tensor parallelism & Model \newline(spatial) & Frequent fine-grained collectives, per layer per token & Mostly serving-dependent, as per-token volume is fixed by hidden dimension and TP degree; total volume grows with tokens generated \\
\addlinespace
Pipeline parallelism & Model \newline(spatial) & Activation transfer at stage boundaries, per engine step & Both, as activation size is serving-dependent and burst frequency follows batch occupancy \\
\addlinespace
Expert parallelism & Model \newline(spatial) & Token routing and expert all-to-all, per layer per token & Both, as per-token volume is serving-dependent and expert load is workload-dependent \\
\addlinespace
Replication & None \newline(full copy) & No inference-time sync between replicas, only load balancing & Neither, as there is no inter-replica traffic on the critical path \\
\addlinespace
Prefill-to-decode disaggregation & Lifecycle \newline(temporal) & Bulk KV-cache transfer per request, plus handover coordination & Both, as volume scales with prompt length (workload) and per-token KV footprint (serving) \\
\bottomrule
\end{tabular}
\end{table*}

These dimensions compose: a deployment may apply \TP within a node, \PP across nodes within a site, replication across sites, and, orthogonally, disaggregate prefill from decode within each replica. Table~\ref{tab:distribution-mechanisms} summarises the communication requirements imposed by each mechanism and whether the resulting cost is fixed by the serving configuration or varies with the workload. The present work explores model parallelism (\TP and \PP), replication (\DP), and functional disaggregation (P/D-D) in a geographically distributed \WAN.

\subsection{SLO-Aware LLM Serving}%
\label{sec:background:slo}

As \LLM inference has moved from research prototypes to production services, \SLOs have become a primary means of expressing serving quality from the user's perspective. Unlike traditional request response services, \LLM requests are generated incrementally: a prompt is first processed to produce the initial output token (i.e., Prefill), after which subsequent tokens are generated autoregressively (i.e., Decode). This two phase structure gives rise to distinct latency characteristics and consequently to two commonly used serving metrics. \TTFT measures the elapsed time from request arrival until the first output token is generated, and is strongly influenced by queueing, scheduling, and the compute intensive prefill phase. \TPOT measures the latency between subsequent output tokens and is strongly influenced by the memory bandwidth and batch characteristics of the decode phase~\citep{huang2025sloaware}. An interactive application such as a chatbot is sensitive to both: a high \TTFT is perceived as an unresponsive system, while a high \TPOT reduces the rate at which generated content becomes available to the user. Other workloads may weight these objectives differently or instead impose end to end latency or throughput targets where token level responsiveness is less important, such as batch or offline processing.

An \SLO specifies a target on one or more such indicators, and \emph{\SLO attainment}, the proportion of requests meeting their assigned target, is related to but distinct from \emph{goodput}, the rate of requests completed within their \SLO per unit time~\citep{wang2024revisiting}: maximising throughput or minimising average latency does not necessarily maximise the number of requests that satisfy their individual \SLOs, particularly once queueing is taken into account. This distinction is sharpened under heterogeneous \SLOs, where a single deployment serves multiple applications or service classes concurrently, each with different \TTFT, \TPOT, or latency requirements, and a request with a relaxed target can absorb delay that would cause a stricter request to violate its own; recent \SLO-aware serving systems therefore treat \SLO requirements as explicit inputs to scheduling rather than treating all requests as interchangeable~\citep{hong2025sola,tang2025scorpio}. For geographically distributed inference, this introduces a further dimension: \emph{where} a request is served becomes part of the scheduling decision, since network latency, queueing conditions, and available capacity all vary by site. For InferSlice, \SLO-awareness therefore provides the basis for treating a geographically distributed infrastructure as a shared resource pool, where placement is driven not only by proximity or utilisation but by which candidate site can satisfy a request's \SLO.

\subsection{Telco Infrastructure as an AI Platform}%
\label{sec:background:telco}

Telecommunications networks provide a mature foundation for differentiated service delivery. Network slicing enables multiple logical services to share physical infrastructure while receiving differentiated resource allocations and performance objectives~\citep{alam2025oran}. However, these objectives have historically focused primarily on network-level characteristics such as bandwidth, latency, reliability, and isolation, rather than the application-level compute requirements of AI services. Liu et al. extend this principle directly to \LLM workloads through LLM-Slice, which creates dedicated wireless network slices for \LLM services and binds communication resources to service requests~\citep{liu2024llmslice}. This demonstrates how service differentiation can be applied to the network delivery of \LLM services, although the work focuses on communication resource control rather than the placement and orchestration of inference computation. The evolution toward \MEC further extended the role of telcos from connectivity toward distributed compute, placing resources closer to users and network access points~\citep{taleb2017mec}. While the broader \MEC vision faced challenges including fragmented infrastructure and limited demand for generic edge workloads, the emergence of telco cloud and AI-RAN, wherein AI and RAN workloads can run concurrently on shared, GPU-accelerated infrastructure, is renewing interest in operator infrastructure as a platform for AI~\citep{kundu2025airan}. AI-RAN primarily targets the co-location of AI and RAN workloads at the radio edge, with its broader commercial opportunity depending on the emergence of latency-critical, edge-native applications such as physical AI, robotics, and autonomous systems.

\LLM inference presents a complementary opportunity that does not require compute to be placed at the radio edge. Its service quality depends jointly on compute capacity, network conditions, queueing, and workload characteristics, while the geographical distribution and operational control of telco infrastructure can provide properties that are difficult to obtain from centralised cloud platforms: data and service sovereignty, stronger security and governance, lower latency through proximity to users, and greater predictability and reliability through controlled network and compute resources. This creates an opportunity to use existing telco compute sites, including central offices, regional data centres, and core network locations, as a distributed substrate for AI serving. In this setting, the service-differentiation principles long applied to network connectivity can be extended to AI services themselves, with application-level \SLOs guiding the allocation and placement of distributed inference resources.

\subsection{Related Work}%
\label{sec:background:relatedwork}

Recent production systems have demonstrated increasingly sophisticated approaches to \LLM serving under heterogeneous workloads and \SLO requirements. Meta's large-scale study of Llama inference explores millions of deployment configurations spanning heterogeneous accelerators, tensor, pipeline, expert, context, and data parallelism, as well as continuous batching and prefill-decode disaggregation, to maximise throughput while meeting latency \SLOs~\citep{cho2026optimizing}. Similarly, Mitra et al. systematically evaluate disaggregated inference across hundreds of thousands of configurations, showing that its benefits depend strongly on workload characteristics, model size, dynamic rate matching, and elastic scaling~\citep{mitra2026beyond}. Behdin et al. further demonstrate the scale and workload-specific optimisation required for industrial \LLM serving, achieving substantial throughput improvements through the joint optimisation of models and serving infrastructure~\citep{behdin2026scaling}. These works demonstrate the importance of jointly considering inference configuration, workload characteristics, and \SLOs, but primarily target tightly coupled compute infrastructures. Meta's study is particularly representative of this setting: although it considers diverse hardware and deployment configurations, the optimisation problem is formulated within the infrastructure available to a large-scale serving platform rather than across geographically distributed telco sites~\citep{cho2026optimizing}.

More closely related to telecommunications, Yet et al. evaluate SLA-aware \LLM and VLM inference across device, RAN-edge, and cloud tiers using a 5G Standalone AI-RAN testbed~\citep{yet2026ai_on_ran_infocom_workshop}. Their experiments demonstrate the impact of execution location, model configuration, queueing, and network delay on SLA attainment, but use a fixed baseline policy and a hierarchical device-to-RAN-to-cloud topology rather than a distributed pool of independently provisioned serving sites. Wang et al. propose a \emph{wide-area distributed inference network} that separates prefill at edge nodes from decode in a central cloud, with the two stages coordinated over a wide-area network~\citep{wang2026pd}. This work demonstrates the feasibility of \LLM inference across wide-area telco infrastructure and is particularly relevant to the present setting, but its deployment follows a predetermined functional placement rather than dynamically selecting among multiple sites and inference configurations according to application-level \SLOs. GORGO further considers network-aware \LLM serving across geographically distributed inference replicas, jointly accounting for network latency, queueing, and \KV-cache locality when routing requests to minimise \TTFT and end-to-end latency~\citep{gorgo2026}. While this work demonstrates the importance of network conditions in cross-region inference decisions, it operates primarily at the request-routing layer and assumes an existing set of serving replicas, rather than jointly orchestrating compute resources, inference configurations, serving locations, and network resources across distributed sites.

Petals demonstrates that very large LLMs can be executed across Internet-connected, heterogeneous, volunteer resources using fault-tolerant layer partitioning, routing, and dynamic load balancing across unreliable peers~\citep{borzunov2023distributed}. While this work establishes the feasibility of geographically distributed inference over the public Internet, its objective is collaborative execution across unmanaged resources. In contrast, InferSlice considers operator-controlled telco infrastructure and investigates how network-aware service differentiation, resource slicing, and SLO-driven placement interact with multiple inference strategies (TP, PP, DP and P/D-D) over a WAN. To the best of our knowledge, there is no existing reference architecture that brings these dimensions together by treating geographically distributed telco compute sites as a shared substrate for \SLO-driven \LLM inference, with serving location and inference configuration jointly selected according to application requirements, site-level resource availability, and network conditions. This motivates the reference architecture developed in this work.

\section{System Design}%
\label{sec:design}

\begin{figure*}
    \centering
    \includegraphics[width=\linewidth]{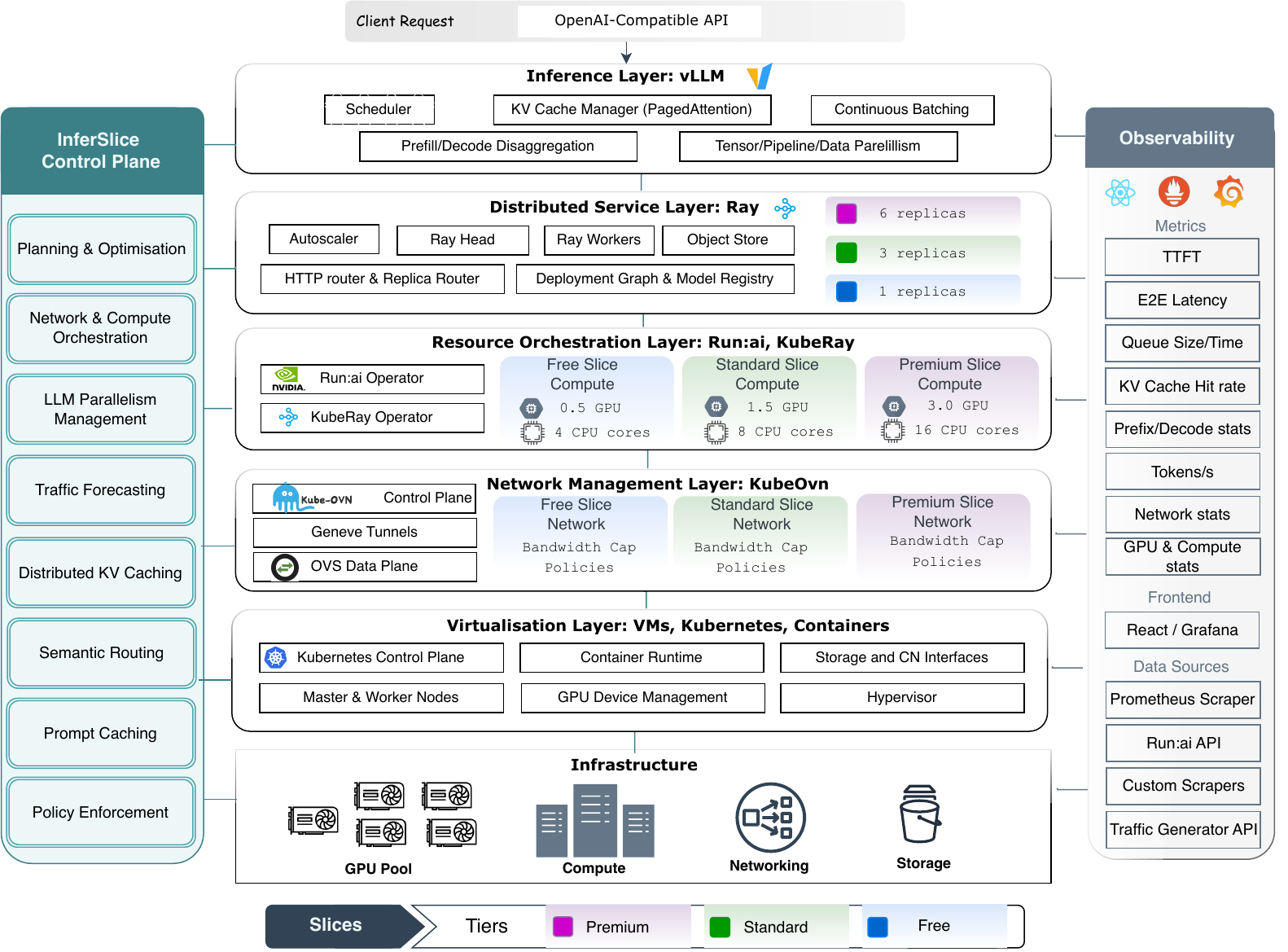}
    \caption{InferSlice reference architecture. Geographically distributed edge
\PoPs provide GPU-accelerated compute interconnected over a \WAN, and are
virtualised and containerised to give isolated, portable execution environments.
Network slicing maps each service tier (Free, Standard, Premium) to its
own logical network with differentiated QoS, while compute orchestration assigns
each tier a guaranteed share of GPU and CPU capacity. A distributed serving layer
expresses workloads as models and replicas and scales them within each tier's
resource envelope, and an inference optimisation layer manages model execution on
the accelerator, including batching and \KV-cache management. A control plane for
planning, traffic forecasting, semantic routing, and prompt and \KV-cache reuse
configures these layers against per-tier SLOs, using cross-layer telemetry from
the observability layer.}
\label{fig:architecture}
\end{figure*}

\subsection{Physical Infrastructure Layer}%
\label{sec:design:infra}

For the development of InferSlice, we assume a telecom-native infrastructure
as a set of geographically distributed edge computing sites,
as discussed in Section~\ref{sec:background:telco}.
The physical infrastructure layer, represented at the bottom of Figure~\ref{fig:architecture},
consists of a set of compute hosts, each providing CPU, memory, storage,
and hardware acceleration, both GPU and \NIC.
These hosts are interconnected by a \WAN and together form edge computing \PoPs,
which allow InferSlice to optimise the placement of the inference workload and
reduce latency and network overhead
(for example, serving a user from the \PoP nearest to them).

In the deployed system, the physical layer is modelled after the \acs{JOINER}
federated testbed where each \PoP comprises Dell R660/R760 host machines
with up to 64 physical CPU cores,
512~GB of RAM and an NVIDIA L40S GPU accelerator.
Each host's \NIC is connected directly to the \PoP's edge router (Juniper ACX7100) via 10~Gbps SFP optics.
Since the \JOINER platform itself is multi-tenant,
compute resources are exposed as \VMs to InferSlice.
InferSlice's storage is based on the general-purpose disks of the compute nodes,
which the virtualisation layer pools into shared volumes.
Further details of the testbed, including the host hardware each node carries,
are given in the Annex~\ref{sec:annex-joiner}.

\subsection{Virtualisation Layer}%
\label{sec:design:virt}
To enable scalability and configurability for modern AI workloads, a virtualisation layer is necessary. The resources of the testbed's tenants are isolated from one another using
\JOINER's virtualisation capabilities, implemented on top of VMware's hypervisor and automation platform.
Similarly, the physical network connectivity is abstracted
as a \LtwoVPN underlay implemented by the testbed's \SDWAN,
with \MTU configured to 9000 to support the so-called jumbo frames and reduce the network overhead.
As a result, the \PoPs attach to a mesh network, exposed as a single ``logical" data-link.
In addition to the base virtualisation, InferSlice further subdivides compute
and network resources by employing a uniform containerised execution environment.
Each inference service is packaged as a small group of container images,
designed to be self-sufficient and to be deployed easily on any VM.
Since containerisation bundles the required libraries into the image itself,
and any auxiliary services (such as databases and proxy-servers) are deployed as supporting containers,
this technique avoids dependency conflicts.
Moreover, because containers share the host kernel, they are a lightweight form of virtualisation:
workloads deploy faster, and scaling out is easily achievable by starting more copies of a container.

The containerisation is achieved by implementing a Kubernetes cluster.
We use \RKE across all \PoPs in the testbed.
Containers run on the \texttt{containerd} runtime and GPU acceleration is exposed to
the scheduler by the NVIDIA GPU Operator together with the NVIDIA Container Toolkit
\footnote{\url{https://docs.nvidia.com/datacenter/cloud-native/container-toolkit/latest/index.html}}.
This means that each L40S GPU accelerator exposed to the platform can be shared among multiple containers.
Persistent storage is provided by two complementary technologies built on top of the
underlying node storage: Longhorn\footnote{\url{https://longhorn.io/}} supplies block
volumes for caches and application data, while model weights can optionally be served
by on-demand streaming from MinIO\footnote{\url{https://www.min.io/}} object storage.

\subsection{Network Isolation and Slicing Layer}%
\label{sec:design:network}

Network slicing logically partitions a shared physical network into multiple isolated networks~(i.e., network slices), each of which can be associated with its own connectivity, resource allocation, and \QoS policies~\cite{wu2025drl}.
Such isolated network slices enable service differentiation by allowing network and compute resources to be allocated and enforced independently across service tiers.
This capability is particularly important for telecom operators deploying LLM inference over existing distributed infrastructure: heterogeneous inference services with different latency and throughput requirements~\cite{wu2025service} must coexist across the same PoPs and over the shared WAN fabric, while retaining differentiated \QoS performance.
Rather than provisioning separate physical network infrastructure for each inference service with a specific \QoS requirement, slicing allows an operator to expose differentiated service tiers over a common infrastructure with service-specific \QoS differentiations in a dynamic, on-demand manner.
In InferSlice, we adopt this operator-oriented abstraction to couple network-level service differentiation with the corresponding compute resources assigned to each inference tier.

Network isolation is implemented using Kube-OVN~\cite{kubeovn}, an OVN-based \CNI plugin for Kubernetes.
InferSlice maps each service tier---\textit{Free}, \textit{Standard}, and \textit{Premium}---to a separate logical network~(a subnet).
This provides each tier with an independent network segment over which connectivity and \QoS policies can be enforced.
In our implementation, per-tier bandwidth limits and traffic policies are enforced at the OVS dataplane, allowing network resources to be differentiated without requiring physically separate network infrastructure.

\begin{figure}[t]
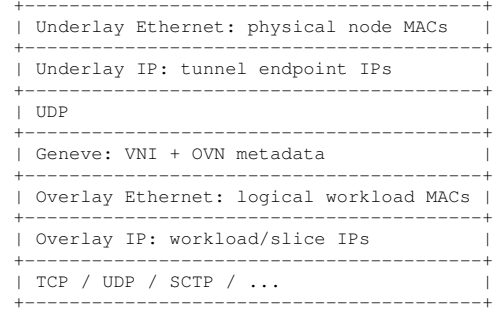

\centering
\begin{minipage}{0.98\columnwidth}
\scriptsize
\begin{verbatim}
    +-----------------------------------------+
    | Underlay Ethernet: physical node MACs   |
    +-----------------------------------------+
    | Underlay IP: tunnel endpoint IPs        |
    +-----------------------------------------+
    | UDP                                     |
    +-----------------------------------------+
    | Geneve: VNI + OVN metadata              |
    +-----------------------------------------+
    | Overlay Ethernet: logical workload MACs |
    +-----------------------------------------+
    | Overlay IP: workload/slice IPs          |
    +-----------------------------------------+
    | TCP / UDP / SCTP / ...                  |
    +-----------------------------------------+
\end{verbatim}
\end{minipage}
\caption{Simplified Geneve packet encapsulation in the Kube-OVN
overlay network.}
\label{fig:geneve-encapsulation}
\end{figure}

Connectivity between workloads belonging to the same logical network is provided by the Kube-OVN overlay network.
When communicating workloads reside on different nodes, Kube-OVN encapsulates the logical network packets using Geneve tunnels and transports them over the physical underlay network~(the JOINER infrastructure network).
The Geneve encapsulation preserves the original Ethernet and IP headers of the logical network while adding the tunnel endpoint addresses required for communication between the corresponding Kubernetes nodes.
The preservation of the inner network headers is particularly useful for \QoS enforcement in our architecture.
Since each isolated network slice is associated with a distinct IP address space~(i.e., subnet), packets can be classified according to their inner source and destination addresses and mapped to the corresponding slice \QoS policy.
A simplified Geneve packet encapsulation in our architecture is illustrated in Fig.~\ref{fig:geneve-encapsulation}.

This layer ensures that traffic from different service tiers is isolated at the network level, preventing cross-tenant interference and enabling per-tier \QoS guarantees. The shared JOINER subnet represents a current contention point under high-load scenarios, which we discuss in Section~\ref{sec:discussion}.

\subsection{Compute Orchestration and Resource Scheduling Layer}%
\label{sec:design:orchestration}

The compute orchestration and resource scheduling layer manages how the shared accelerator pool is partitioned and allocated across service tiers and distributed inference workloads. Network isolation alone cannot guarantee service differentiation, since inference \QoS ultimately depends on the compute capacity available to each workload. The layer therefore provides compute isolation and resource guarantees by assigning each tier a defined share of GPU and CPU capacity, together with quotas and scheduling priorities, while allowing resources to be shared across tiers. This is particularly important for heterogeneous service requirements: lower-priority workloads may benefit from fractional GPU allocation, whereas higher-priority tiers require sufficient reserved capacity to maintain their SLOs under contention. Together with the network management layer, the compute layer establishes a coordinated resource envelope in which GPU, CPU and network capacity are allocated according to the requirements of each service tier.

In InferSlice, this functionality is implemented using NVIDIA Run:ai~\cite{runai} and KubeRay~\cite{kuberay}, which provide complementary resource management and distributed workload orchestration capabilities within Kubernetes. Each service tier is mapped to a Run:ai project defining its GPU and CPU resource allocations, quotas and scheduling priorities, as shown in Fig.~\ref{fig:architecture}. Run:ai schedules workloads against these resource constraints and supports fractional GPU allocation, allowing multiple inference workloads to share the available accelerator pool while enforcing the resource boundaries of each tier. This enables, for example, the Free, Standard and Premium slices to use different GPU shares without requiring dedicated hardware for each tier.

KubeRay manages the lifecycle of the distributed inference deployments that consume these resources. Declarative Ray-based deployment specifications are reconciled into the corresponding Kubernetes pods, which are subsequently scheduled within the compute resources assigned to their service tier. Replica counts are constrained by the available slice capacity, preventing scaling in one tier from consuming resources guaranteed to another. GPU and compute utilisation, scheduling state and workload status are exposed through the observability layer, providing the measurements required by the InferSlice control plane to monitor resource availability and, ultimately, support higher-level planning, optimisation and SLO-aware orchestration.

\subsection{Distributed Serving and Replica Management Layer}%
\label{sec:design:serving}

The distributed serving and replica management layer acts as middleware between the inference workload and the virtualised resources managed beneath it.
Where the orchestration layer manages and schedules resources, this layer exposes them to the application through a distributed computing abstraction, so that the workload is expressed in application terms (e.g., models, replicas, and per-replica resource requirements) rather than in terms of nodes, devices, and pods.
It translates these application-level requirements into resource requests the orchestration layer can satisfy and coordinates the resulting distributed components at runtime.
This separation is needed because inference workloads must scale and recover independently of where their resources are physically located.
Without such a layer, each workload would need to be tied to specific infrastructure, preventing the same resource pool from being shared across tiers and applications.
The layer therefore provides a distributed runtime for the inference workload and manages the number and lifecycle of its replicas, while remaining agnostic to the underlying infrastructure.

In InferSlice, this functionality is implemented using Ray~\cite{moritz2018ray}, deployed on Kubernetes through KubeRay as described in Section~\ref{sec:design:orchestration}.
Ray provides a distributed runtime in which a head node coordinates the cluster, worker nodes host the workload, and a shared object store supports data exchange between distributed components, as shown in Fig.~\ref{fig:architecture}.
Each model deployment is described in a deployment graph and tracked in a model registry, together with the resources each of its replicas requires.
These application-level specifications are mapped to resource requests submitted to the underlying Kubernetes and Run:ai layers, which place the corresponding pods within the compute slice of the tier.
The Ray autoscaler adjusts the number of replicas of each deployment in response to serving demand, subject to minimum and maximum bounds and to the capacity of the tier's compute slice, and issues the corresponding resource requests when replicas are added or removed.

\subsection{Inference Optimisation Layer}%
\label{sec:design:inference}

The inference optimisation layer manages the lifecycle and execution of the model within each serving replica, from model loading and initialisation to GPU-resident execution and exposure through the inference API.
Whereas the distributed serving layer manages replicas as application-level units, this layer determines how the model is instantiated and operated within each replica, including model loading, GPU memory allocation, request scheduling, batching, and \KV-cache management.
It therefore provides the abstraction between a model deployment and the underlying accelerator, hiding device-level execution details from the serving and orchestration layers.
This separation is important because the same model can require different execution strategies depending on available GPU memory, workload characteristics, concurrency, and SLO requirements.
The layer converts a model from a deployment artefact into a running inference service while exposing a consistent interface to the distributed serving layer.

In InferSlice, this functionality is implemented using vLLM~\cite{kwon2023vllm} as the inference runtime.
For each replica, vLLM loads the model weights, initialises the model on the allocated GPU resources, manages GPU memory during execution, and exposes an OpenAI-compatible inference interface to the serving layer.
The runtime scheduler coordinates concurrent requests through continuous batching, while the \KV-cache manager allocates and reuses attention states during autoregressive generation.
These mechanisms allow multiple requests to share the resources of a replica while efficiently managing GPU memory and compute across concurrent requests.
The layer also provides the mechanisms required for distributed inference, including tensor, pipeline, and data parallelism and prefill/decode disaggregation~\cite{zhong2024distserve}, allowing the model to be partitioned or replicated according to the configuration specified in its deployment.

\subsection{Control Plane}%
\label{sec:design:control}

The InferSlice control plane provides the decision-making layer that coordinates compute, network, serving, and inference resources to meet the SLOs of each service tier.
Whereas the underlying layers execute resource allocation, replica management, and model inference, the control plane determines how these resources and execution mechanisms should be configured and where workloads should be served.
It maintains a global view of workload demand, service requirements, resource availability, network conditions, and observed system performance, and uses this information to determine the serving configuration for each tier.
This separation is essential in a distributed operator environment, where compute and network resources span multiple sites and the configuration that satisfies an SLO can change as demand and infrastructure conditions evolve.
The control plane therefore provides the coordination mechanism through which local resource management is transformed into end-to-end, SLO-aware service orchestration.

In InferSlice, the control plane comprises planning and optimisation, traffic forecasting, semantic routing, prompt caching, and distributed \KV-cache management functions, as shown in Fig.~\ref{fig:architecture}.
The planning and optimisation component determines the required serving configuration based on the SLOs and available resources, including replica counts, serving locations, resource allocations, and inference configurations.
Traffic forecasting provides estimates of future demand to support proactive scaling and placement decisions, while semantic routing directs requests towards replicas according to application and service requirements.
Prompt caching and distributed \KV-cache management further exploit reuse across requests and replicas, reducing redundant computation and the cost of transferring state across the network.
Together, these components provide the control logic required to coordinate the underlying compute, network, serving, and inference layers.

The control plane operates over the state exposed by the observability layer and translates this state into configuration changes applied to the underlying infrastructure.
Resource utilisation, request rates, queueing behaviour, latency, throughput, network conditions, replica state, and \KV-cache residency provide the feedback required to identify SLO violations and available capacity.
The resulting decisions are then enforced through the compute orchestration, network slicing, distributed serving, and inference optimisation layers, closing the loop between observed system state and service configuration.
This architecture separates policy and optimisation from resource enforcement, allowing the same control plane to reason over heterogeneous infrastructure and different inference configurations while leaving the underlying layers responsible for their respective execution mechanisms.

\subsection{Observability Layer}%
\label{sec:design:observability}

The observability layer provides the measurements required to understand the state and performance of the distributed inference system across all infrastructure and service layers.
Whereas the control plane determines how the system should be configured, the observability layer provides the telemetry required to make those decisions based on current and historical system state.
It therefore collects and aggregates metrics across compute, network, serving, and inference components, providing a consistent view of resource utilisation, workload behaviour, and SLO attainment.
This cross-layer visibility is particularly important in a distributed environment, where performance degradation may originate from compute contention, network congestion, replica saturation, or inference-level effects such as \KV-cache pressure.
The layer consequently provides the measurement and feedback mechanism that connects the execution layers with the control plane.

In InferSlice, observability spans resource, network, serving,
and inference telemetry,
including GPU and CPU utilisation, allocated resources,
network throughput and latency,
request rates, queueing, \TTFT, \TPOT,
token throughput, replica state,
and \KV-cache utilisation and hit rate,
as shown in Fig.~\ref{fig:architecture}.
Telemetry is gathered by a combination of standard and custom collectors.
Prometheus scrapers collect metrics exposed by the Kubernetes, serving, and inference components, the Run:ai API provides GPU allocation and scheduling state for each tier, and custom scrapers collect measurements not available from standard exporters, such as network statistics and the request-level metrics reported by the traffic generator.
Prometheus stores and aggregates these measurements as time series, and Grafana and a React frontend provide the interfaces through which they are queried and visualised. InferSlice's React frontend is described in the Appendix Section~\ref{sec:dashboard}.

The resulting telemetry supports both runtime monitoring and offline evaluation: current measurements allow resource pressure and SLO violations to be identified, while historical measurements support traffic forecasting, capacity planning, and analysis of the operating boundaries of different inference configurations.
A common observation layer across geographically distributed resources allows InferSlice to analyse end-to-end service behaviour, rather than treating compute, network, and inference performance as independent systems.

\section{Experimental Setup}%
\label{sec:setup}

\subsection{Testbed and Workload}%
\label{sec:setup:testbed}

To assess \SLO-driven inference, we deployed InferSlice across a representative subset of six \PoPs from the \JOINER federated testbed, selected to capture variation in inter-\PoP distance.
Table~\ref{tab:testbed-vms} lists the \VMs at each site; six of them carry a
single NVIDIA L40S accelerator, attached by PCI passthrough, with 48~GB of GPU
memory and 864~GB/s of memory bandwidth.
The \VMs run Ubuntu 24.04 LTS on Intel Xeon Gold 6438Y+ hosts,
provisioned with 16 vCPUs,
62~GiB of RAM and roughly 500~GB of local disk,
and running the software stack summarised in Table~\ref{tab:testbed-software}.
Each service tier is served by a set of vLLM replicas deployed
as Ray services and orchestrated by KubeRay (Section~\ref{sec:design:orchestration}),
and accessed through vLLM's standard OpenAI-compatible interface.
In turn, each replica is configured to reserve up to 92\% of the GPU memory granted to it by Run:ai,
which is shared between the model and its KV cache.
Model weights are staged in a distributed MinIO pool and streamed to replicas on demand.
\VMs without a GPU host the cluster's control plane and auxiliary functions,
such as monitoring, API servers, persistent storage and reverse proxying,
as well as traffic generators for collecting experimental performance data.
The traffic exchanged between distributed replicas (for example, \KV-cache and
collective transfers across \PoP nodes) is carried over the standard TCP/IP network.

\begin{table}[t]
\centering
\caption{\PoP and computing capabilities used in the evaluation}
\label{tab:testbed-vms}
\vspace{0.5ex}
\begin{tabular}{lll}
\toprule
\PoP & VM & GPU \\
\toprule
\multirow{2}{*}{Bristol}     & VM1 & --- \\
                             & VM2 & --- \\
\cmidrule(lr){1-3}
\multirow{2}{*}{Southampton (Soton)} & VM1 & L40S \\
                             & VM2 & L40S \\
\cmidrule(lr){1-3}
\multirow{2}{*}{Oxford}      & VM1 & --- \\
                             & VM2 & L40S \\
\cmidrule(lr){1-3}
\multirow{2}{*}{London (UCL)}         & VM1 & --- \\
                             & VM2 & L40S \\
\cmidrule(lr){1-3}
\multirow{2}{*}{Belfast (QUB)}         & VM1 & --- \\
                             & VM2 & L40S \\
\cmidrule(lr){1-3}
Dublin (TCD)                          & VM1 & L40S \\
\bottomrule
\end{tabular}
\end{table}

\begin{table}[t]
\centering
\caption{Software stack supporting LLM inferece.}
\label{tab:testbed-software}
\vspace{0.5ex}
\begin{tabular}{ll}
\toprule
Component & Version \\
\midrule
Kubernetes (RKE2)      & 1.34.3 \\
Container runtime (containerd) & 2.1.5 \\
Kube-OVN               & 1.14.15 \\
NVIDIA GPU Operator    & 25.10.1 \\
NVIDIA driver          & 580.126.09 \\
CUDA                   & 13.0 \\
NVIDIA Container Toolkit & 1.18.1 \\
NCCL                   & 2.27.5 \\
vLLM                   & 0.12.0 \\
Ray                   & 2.53.0 \\
KubeRay                   & 1.2.2 \\
Longhorn               & 1.10.1 \\
MinIO server           & R.2025-04-08 \\
MinIO operator         & 7.1.1 \\
Run:ai                 & 2.24.11 \\
\bottomrule
\end{tabular}
\end{table}

Client requests are received through a standard OpenAI-compatible interface.
To perform the initial functional validation and to emulate the workload,
we use the vLLM benchmark CLI which sends independent,
short-lived, stateless chat requests drawn from the ShareGPT dataset with a
Poisson arrival process.
The tool supports this dataset natively, and the Poisson process gives
fine-grained control over the arrival pattern~\citep{vllmBench}.
For the quantitative experiments, we set the arrival rate to 100 requests/s
and employed a varying number of parallel ongoing requests between 2 and 32.

The evaluation comprises three types of experiments:
a service-tier experiment,
which varies load and concurrency across the Free,
Standard and Premium tiers~(Section~\ref{sec:exp_servicediff}), elastic scale-out effects and optimisation (Section~\ref{sec:exp_cold_start}),
and a distributed-inference experiment,
in which the model is partitioned across \PoP nodes~(Section~\ref{sec:distributed_ai}).
We served each type with a different model.
\texttt{Llama-3.1-8B-Instruct} is a dense 8B-parameter,
instruction-tuned conversational model;
since it fits entirely on a single L40S without need for partitioning,
we used it for the service-tier experiments.
\texttt{GPT-OSS-20B}, on the other hand,
is a sparse mixture-of-experts model of 21B parameters (3.6B active) with a 128k context,
designed for low-latency reasoning and agentic workloads;
when its full 21B parameters are loaded into memory,
it is a good candidate for the distributed-inference experiments.

Throughout the experiments, we recorded the following metrics:
\begin{itemize}
    \item \TTFT, the elapsed time from request arrival to the generation of the first output token;
    \item \TPOT, the latency between consecutive generated output tokens;
    \item aggregate token throughput, the total number of tokens generated per second across all requests;
    \item goodput, the rate in requests per second of requests that meet their \TTFT \SLO (here \TTFT $<$ 500\,ms; the distributed-inference experiments of Section~\ref{sec:evaluation:network} relax this to 2\,s, where per-request cross-\PoP costs make the stricter threshold uninformative);
    \item success rate, the proportion of requests completed without error.
\end{itemize}

\subsection{Network Substrate Characterisation}
\label{sec:setup:network}

With Inferslice deployed over the JOINER network, we carried out a network characterisation experiment.
The~$80$ cross-\PoP directed pairs measured reveal a four-tier latency structure that reflects the geographic distribution of the JOINER testbed, as summarised in Table~\ref{tab:network_profile}:

\textbf{Near-adjacent \PoPs} ($<5$~ms, $n=8$): 
Near-adjacent \PoPs~(e.g., pairs between the \texttt{10.203.41.0/24} and \texttt{10.203.43.0/24} subnets) exhibit ICMP latencies of approximately $4.06$~ms--$4.51$~ms with TCP throughput of $6.47$~Gbps--$6.59$~Gbps~(mean $6.53$~Gbps).
These PoP pairs are plausible candidates for inter-\PoP pipeline parallelism, as the network overhead is commensurate with intra-server NVLink latency at scale.

\textbf{Mid cross-\PoPs} ($5$~ms--$10$~ms, $n=32$):
Mid cross-\PoPs~(e.g., pairs between the \texttt{10.203.43.0/24} and \texttt{10.203.45.0/24} subnets) exhibit a mean ICMP latency of approximately $7.57$~ms and mean TCP throughput of $2.79$~Gbps.
These paths introduce moderate communication overhead but remain suitable candidates for pipeline parallelism with sufficiently coarse-grained inter-stage communication.

\textbf{Far cross-\PoPs} ($10$~ms--$15$~ms, $n=8$):
Far cross-\PoPs~(e.g., pairs between the \texttt{10.203.44.0/24} and \texttt{10.203.45.0/24} subnets) exhibit a mean ICMP latency of approximately $10.90$~ms and mean TCP throughput of $1.19$~Gbps.
The increased latency and reduced throughput impose greater communication overhead, making these paths less suitable for tightly coupled synchronous operations and favouring pipeline configurations with less frequent inter-stage communication.

\textbf{Distant \PoPs} ($\geq15$~ms, $n=32$):
Distant \PoPs~(e.g., pairs between the \texttt{10.203.42.0/24} and \texttt{10.203.43.0/24} subnets) exhibit ICMP latencies of approximately $15$~ms--$22$~ms, with a mean of $18.97$~ms, and mean TCP throughput of $1.30$~Gbps.
These paths are unsuitable for synchronous parallelism strategies and motivate asynchronous or speculative execution models for cross-\PoP inference disaggregation.

In addition to the cross-\PoP connections, there are~$10$ directed intra-\PoP pairs~(i.e., connections within the same JOINER \PoP site), which exhibit a mean ICMP latency of approximately $0.60$~ms and mean TCP throughput of approximately $10.0$~Gbps.



\begin{table*}[h]
\centering
\caption{JOINER testbed network characterisation summary across
         90 directed node pairs, comprising 80 cross-\PoP and
         10 intra-\PoP pairs. TCP throughput values are
         unidirectional (TX).}
\label{tab:network_profile}
\begin{tabular}{lccc}
\toprule
\textbf{Latency tier} & \textbf{Pairs} &
\textbf{Mean ICMP (ms)} & \textbf{Mean TCP TX (Gbps)} \\
\midrule
Near cross-PoP ($<5$~ms)          & $8$  & $4.26$  & $6.53$ \\
Mid cross-PoP ($5$~ms--$10$~ms)   & $32$ & $7.57$  & $2.79$ \\
Far cross-PoP ($10$~ms--$15$~ms)  & $8$  & $10.90$ & $1.19$ \\
Distant cross-PoP ($\geq15$~ms)   & $32$ & $18.97$ & $1.30$ \\
\midrule
Intra-PoP (same subnet)            & $10$ & $0.60$  & $10.0$ \\
\bottomrule
\end{tabular}
\end{table*}

This four-tier cross-\PoP structure has direct implications for InferSlice's placement policy: tensor parallelism should be confined to intra-\PoP pairs; pipeline parallelism is viable up to approximately $10$~ms; and expert parallelism for mixture-of-experts models, which tolerates higher latency due to conditional routing sparsity, is the preferred parallelism strategy for distant cross-\PoP pairs.
Section~\ref{sec:discussion} returns to these implications in the context of the inference results below.

\subsection{Experiments Definition}

\subsubsection{Service Differentiation}
\label{sec:exp_servicediff}
\begin{figure}
    \centering
    \includegraphics[width=\linewidth]{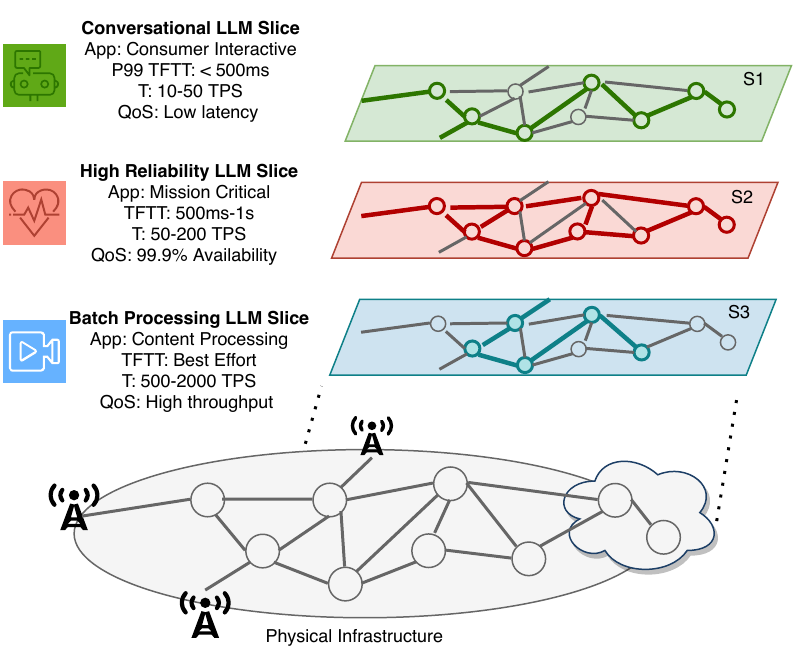}
    \caption{Three LLM service slices over shared telco infrastructure,
    each enforcing distinct latency, throughput, and availability
    policies without dedicated hardware per tenant.}
    \label{fig:slice_tiers}
\end{figure}

Telco \LLM deployments must support a heterogeneous mix of AI applications with fundamentally different \SLOs. Conversational agents and voice assistants require low and predictable latency, with interactive \LLM serving commonly targeting P99 \TTFT below 500\,ms and bounded token generation latency to maintain acceptable user experience and conversational flow~\cite{mlperf_v51}. In contrast, network operations assistants analysing multimodal telemetry may prioritise service availability, reliability, and predictable response times, while large-scale workloads such as content moderation, document processing, or synthetic data generation are primarily constrained by throughput and cost efficiency. These workload classes impose conflicting requirements that cannot be optimally satisfied by a single inference operating point. Consequently, resource allocation, request admission, autoscaling behaviour, and network resource assignment must be differentiated according to application \SLOs. Figure~\ref{fig:slice_tiers} illustrates this concept through three representative \LLM service slices, each enforcing distinct \QoS objectives while sharing the same underlying infrastructure.

To evaluate whether InferSlice can deliver differentiated inference services over shared telco infrastructure, we conducted a service differentiation experiment on the JOINER platform. Three slices, \emph{Free}, \emph{Standard} and \emph{Premium}, were instantiated over the same physical GPU pool and differ in the resources they are entitled to (Figure~\ref{fig:architecture}). At the compute layer, Run:ai assigns each slice a fractional GPU project of 0.5, 1.5 and 3 GPUs, with 4, 8 and 16 CPU cores respectively, so that prioritisation is enforced without dedicating hardware to a tenant. At the inference layer, KubeRay autoscales each slice up to a ceiling of 1, 3 and 6 replicas, each replica occupying 0.5 GPU. Kube-OVN provides slice-level virtual networking and traffic isolation. All slices serve the same Llama-3.1-8B-Instruct deployment, so any difference in behaviour is attributable to the resources and scaling headroom of the slice rather than to the model or hardware.

Load is generated from the ShareGPT dataset by sweeping the client \texttt{max\_concurrency} from 2 to 32 against each slice. For each setting we record P99 TTFT and SLO-compliant goodput, defined as the rate of requests completing with a TTFT of at most 500\,ms. The goal is to determine at which offered load each slice stops meeting its SLO, and whether the slices degrade in the order implied by their allocations. A second experiment, reported in Appendix~\ref{app:concurrency_policy}, holds the resource allocation fixed and instead differentiates slices through their concurrency policy (\texttt{target\_ongoing\_requests} and \texttt{max\_ongoing\_requests}) under an increasing request rate.

\subsubsection{Cold-start latency}
\label{sec:exp_cold_start}

Elastic scale-out only helps if new replicas become ready to serve
quickly, and for multi-billion-parameter models the time spent acquiring
weights can account for much of the delay between a scaling decision and
the first served response. To reduce this delay, we load model weights
through a parallel streaming reader instead of the default memory-mapped
path. This experiment evaluates that design choice by quantifying how much
cold-start latency parallel streaming removes relative to the default
loader, both when the service is deployed from scratch (cold) and when it
scales back up from its minimum replica count (hot).

To isolate the cost of weight acquisition during elastic scale-out, we
benchmark the premium-tier RayService serving Llama-3.1-8B-Instruct (bf16)
under two weight-loading variants that differ only in the loader. The
default configuration memory-maps safetensors weights from an HF cache
persisted on a Longhorn PVC, whereas the runai-streamer variant loads
the same cache through the Run:ai Streamer parallel filesystem reader
(\texttt{load\_format: runai\_streamer}). Each variant is evaluated in a
cold phase, in which the RayService is deleted and re-applied from
scratch while load is already running, and in a hot phase, in which load
is resumed after the service has scaled back down to its minimum replica
count. A fixed load of 50 concurrent streaming chat completions, sampled
from the conversation dataset used by the production traffic generator
and capped at 256 output tokens per request, maintains per-replica
\texttt{ongoing\_requests} above the autoscaler target and thereby forces
the addition of five replicas under load. Container images are pre-pulled
on all GPU nodes so that image retrieval does not contaminate cold
timings. We record the cold deploy time from manifest apply to serve
readiness and, for each worker pod, the time-to-first-served-response,
identified as the first increment of the pod's
\texttt{ray\_vllm\_request\_success\_total} counter above its phase
baseline in the cluster Prometheus at 1\,s scrape resolution.

\subsubsection{Network bottleneck identification}
\label{sec:distributed_ai}
\paragraph{Experiment N1: End client bandwidth limiting.}
The preceding network characterisation establishes the latency and
throughput available across the JOINER substrate. We next ask when
network capacity, rather than inference compute or request queueing,
becomes the binding constraint on serving performance. We isolate the
client-facing link and apply per-slice bandwidth caps while holding the
serving configuration fixed. The client is located at a fixed site and
the bandwidth cap is varied independently for the Free, Standard, and
Premium slices. For each configuration, we measure P99 \TTFT, P99 \TPOT,
token throughput, and container-level receive and transmit rates.
Experiments are repeated at client concurrency levels of 10, 16, and
32. The 100\,Gbps configuration serves as an effectively unconstrained
baseline; lower caps are selected around the observed traffic demand
to expose the transition from compute- or queue-limited to
network-limited operation.

\paragraph{Experiment N2: Distributed model latency characterisation (TP/PP across \PoP nodes).}
We characterise how model parallelism behaves when the shards of a single replica are placed across \PoP nodes. We fix the model to GPT-OSS-20B and deploy 2-shard tensor-parallel (\texttt{tp2}) and pipeline-parallel (\texttt{pp2}) configurations across measured \PoP pairs at $\sim$9, $\sim$18 and $\sim$33\,ms \RTT. A co-located single-node deployment (\texttt{tp1}) is the baseline, so that any change in behaviour is attributable to shard placement alone. Each arm uses the production serving stack, orchestrated by Ray Serve with a vLLM backend, and streams weights from MinIO; only the parallelism strategy and worker placement differ. Because the parallelism under test sits below the OpenAI API layer, load is generated directly against the vLLM engine with vLLM \texttt{bench serve}. To separate per-request \TTFT from client-side queueing, each arm's offered rate is raised from a low queue-free reference until the median \TTFT departs from the per-request baseline or the P99 develops a queue tail; the highest rate at which the response-time distribution shows no onset of queueing is reported as the configuration's maximum queue-free offered load, and the saturated operating point is reported separately in the text. We report median and P99 \TTFT, output throughput and goodput (requests/s meeting \TTFT$<$2\,s; the 500\,ms service \SLO of Section~\ref{sec:setup:testbed} is relaxed to 2\,s for these WAN experiments, where per-request cross-\PoP costs make the stricter threshold uninformative), and measure per-shard transmit and receive rates from Prometheus byte counters to characterise the inter-shard communication pattern of each strategy. A configuration is deemed feasible if it remains SLO-compliant, which identifies the \RTT range over which each strategy can be distributed.

Four-shard configurations (\texttt{pp4}, \texttt{tp2\_pp2}, \texttt{tp4}, \texttt{tp4\_ep4}) span several inter-node links with different \RTTs, so their results cannot be attributed to a single \RTT value; these configurations do not initialise within a 40-minute deployment budget and are excluded from the comparison.

\paragraph{Experiment N3: Distributed model bandwidth limiting}
With the latency behaviour of each strategy established in Experiment~N2, we hold \RTT fixed at each pair's natural value and sweep the available inter-shard bandwidth downward to find where bandwidth becomes the binding constraint. The deployment follows Experiment~N2, so only the bandwidth between shards changes between runs. Because measured \PoP pairs couple latency and bandwidth, we isolate bandwidth by rate-shaping alone: an HTB rate cap is applied to each shard node's overlay egress towards its peer shard (in both directions, using destination-/32 filters on the Geneve overlay), which leaves the path's \RTT untouched, and each cap is verified on the dataplane before its run. We sweep \texttt{pp2} on three pairs, \augusta--\cyrenaica ($\sim$9\,ms), \augusta--\equestris ($\sim$18\,ms) and \equestris--\claudiaxi ($\sim$33\,ms), and \texttt{tp2} once on \augusta--\cyrenaica. Because \texttt{tp2} sustains only a small fraction of that load, it is swept under a queue-free ten-prompt load as a latency-bound reference. A descending ladder of caps is applied to every pair, with additional coarser caps on the 9\,ms pair, an unconstrained reference repeated after every sweep to detect drift, and a 1\,Gbps null control on every pair. Metrics follow Experiment~N2, augmented with per-shard byte counters to verify on the wire that each cap binds. We define the bandwidth floor as the cap at which median \TTFT doubles relative to the pair's own unconstrained reference.

\paragraph{Experiment N4: Latency characterisation in Functional Disaggregation (P/D-D across \PoP nodes).}
We repeat the setup of Experiment~N2, but focus on functional disaggregation, splitting the prefill and decode phases of inference across the measured \PoP pairs. We fix the model to Llama-3.1-8B-Instruct, which fits on a single node, so that any behaviour change is attributable to the placement of the two phases alone. Each deployment uses one prefill worker and one decode worker, orchestrated by Ray Serve with a vLLM backend, with the \KV-cache transferred between workers over NIXL. The workers are placed across measured \PoP pairs spanning the mid and distant cross-\PoP tiers of Table~\ref{tab:network_profile}, and an aggregated, co-located deployment on a single node is the baseline. Because the cost of disaggregation depends on the volume of \KV-cache moved per request, we use the ShareGPT workload of Section~\ref{sec:setup:testbed} at an offered request rate of 2.7 requests/s with 100 prompts, so that the prompt-length distribution, and hence the transfer size, reflects interactive chat. This rate is the highest queue-free load tested on every placement; the rate was raised in steps until the onset of queueing appeared in the P99 tail. As in Experiment~N2, load is generated with vLLM \texttt{bench serve}, and we report median/P99 \TTFT, \TPOT, output throughput and goodput (requests/s meeting \TTFT$<$2\,s, the relaxed threshold of Experiment~N2), augmented with the measured per-pod \KV-cache transfer volume and per-shard transfer rates. A placement is deemed feasible if it remains \SLO-compliant, which identifies the \RTT range, at each pair's natural bandwidth, over which prefill/decode disaggregation can be distributed for this workload class.
\section{Evaluation}%
\label{sec:evaluation}
\subsection{Service differentiation across slices}
\label{sec:evaluation:tiers}
\begin{figure*}
\centering
\includegraphics[width=\textwidth]{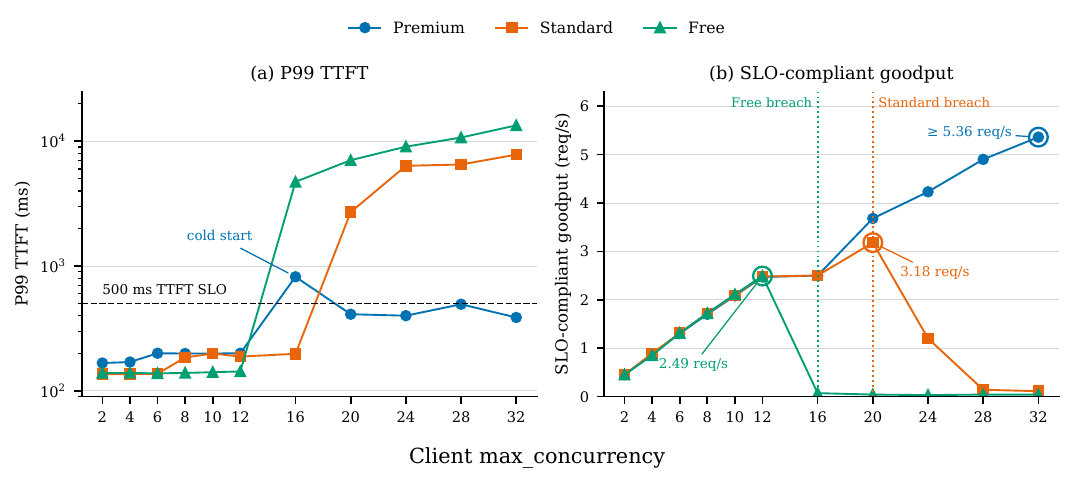}
\caption{Service differentiation across the Free, Standard, and Premium slices (maximum of 1, 3, and 6 replicas) as client \texttt{max\_concurrency} increases. (a) P99 \TTFT against the 500~ms \SLO (log scale). (b) SLO-compliant goodput, the rate of requests whose \TTFT meets the \SLO. Open rings mark each slice's peak and dotted lines mark the concurrency at which Free and Standard first breach the \SLO. The Premium excursion at concurrency 16 is a replica cold start, and its peak is a lower bound because the sweep ends before saturation.}
\label{fig:slice_goodput}
\end{figure*}

We first evaluate whether InferSlice can expose distinct service tiers from a shared infrastructure by differentiating the resources allocated to each slice. The Free, Standard, and Premium slices are allocated a maximum of 1, 3, and 6 replicas respectively. We sweep the client \texttt{max\_concurrency} from 2 to 32 and measure P99 \TTFT and SLO-compliant goodput, defined as the rate of requests whose \TTFT meets the 500~ms \SLO. A complementary experiment in Appendix~\ref{app:concurrency_policy} instead holds the resource allocation aside and differentiates slices through their concurrency policy.

At low load (up to concurrency 12) the three slices are indistinguishable (Figure~\ref{fig:slice_goodput}). P99 \TTFT remains well below the \SLO and goodput grows linearly with load to about 2.5~req/s, because every slice has sufficient capacity for the offered load. The slices diverge once the load exceeds their capacity. Free, limited to a single replica, breaches the \SLO at concurrency 16: P99 \TTFT jumps from about 0.1~s to about 5~s and reaches about 13~s at concurrency 32. Its goodput peaks at 2.49~req/s (462~tok/s) and then collapses to below 0.1~req/s as requests queue behind the saturated replica. Standard, with up to three replicas, sustains the \SLO for longer and breaches at concurrency 20, where P99 \TTFT reaches about 2.8~s and rises to about 8~s at concurrency 32. Its goodput peaks at 3.18~req/s (840~tok/s) at the breach point and falls to about 0.1~req/s at concurrencies 28 and 32. Premium, with up to six replicas, keeps P99 \TTFT below the \SLO across the sweep, apart from a transient excursion at concurrency 16 (about 0.8~s) that we attribute to a replica cold start. Its goodput grows with load to 5.36~req/s (1,026~tok/s) at concurrency 32, where the sweep ends before saturation, so this value is a lower bound.

At their peaks, Premium therefore delivers about 2.2$\times$ the compliant goodput of Free and about 1.7$\times$ that of Standard, while Standard delivers about 1.3$\times$ that of Free. Beyond its capacity a slice does not degrade gracefully: queueing inflates \TTFT for most requests and goodput falls sharply, whereas a slice with sufficient resources continues to scale. These breach points define the compute-limited operating regimes that Experiment~N1 builds on (Section~\ref{sec:distributed_ai}).

\subsection{Cold-Start Reduction}%
\begin{figure}
\centering
\includegraphics[width=0.9\columnwidth]{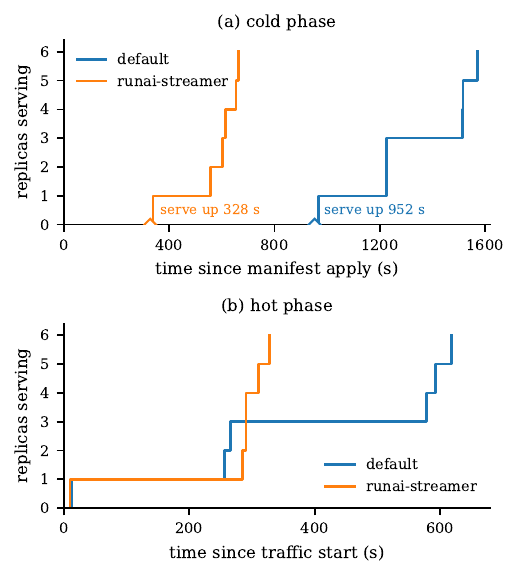}
\caption{Scale-out dynamics of the premium-tier RayService under the
         default and runai-streamer loaders, with a fixed load of 50
         concurrent streaming requests. Each step marks one pod's first
         served response; diamonds mark serve-up. (a)~Cold phase, time
         measured from manifest apply. (b)~Hot phase, time
         measured from traffic start. Hot-phase behaviour matches the
         cold phase for both loaders, indicating that storage-cache
         warmth does not bound weight acquisition.}
\label{fig:coldstart}
\end{figure}

\label{sec:evaluation:coldstart}
During elastic scale-out, weight acquisition dominates
the time required to deploy the pod. With the default loader, a
cold deployment of the premium service takes approximately 950\,s from
manifest apply to serve readiness, and pods added under load serve
their first response after a median of roughly 560\,s, yielding an
end-to-end scale-out time of approximately 1{,}500\,s. Replacing the
default loader with the runai-streamer parallel filesystem reader over
the same PVC cache reduces the cold deploy time to approximately
330\,s, a reduction of about 65\%, and the median per-pod
time-to-first-response to roughly 285\,s, approximately halving the
end-to-end scale-out time to approximately 610\,s. The improvement
cannot be attributed to storage-cache warmth: hot-phase medians match
the cold ones in both variants (roughly 580\,s versus 560\,s with the
default loader, and approximately 290\,s versus 285\,s with
runai-streamer), indicating that the loader, not the cache state,
bounds weight acquisition. Newly added pods in fact reach readiness
within approximately two minutes in both variants, so the observed
differences arise almost entirely during weight loading and
first-token production. These measurements explain the cold-start
transient observed in the service-differentiation sweep
(Section~\ref{sec:evaluation:tiers}) and suggest that weight streaming
can substantially shorten scale-out for latency-sensitive slices.
Figure~\ref{fig:coldstart} visualises these dynamics per pod in the
cold and hot phases.

\subsubsection{Network bottleneck identification}%
\label{sec:evaluation:network}

\begin{figure*}
\centering
\includegraphics[width=0.9\textwidth]{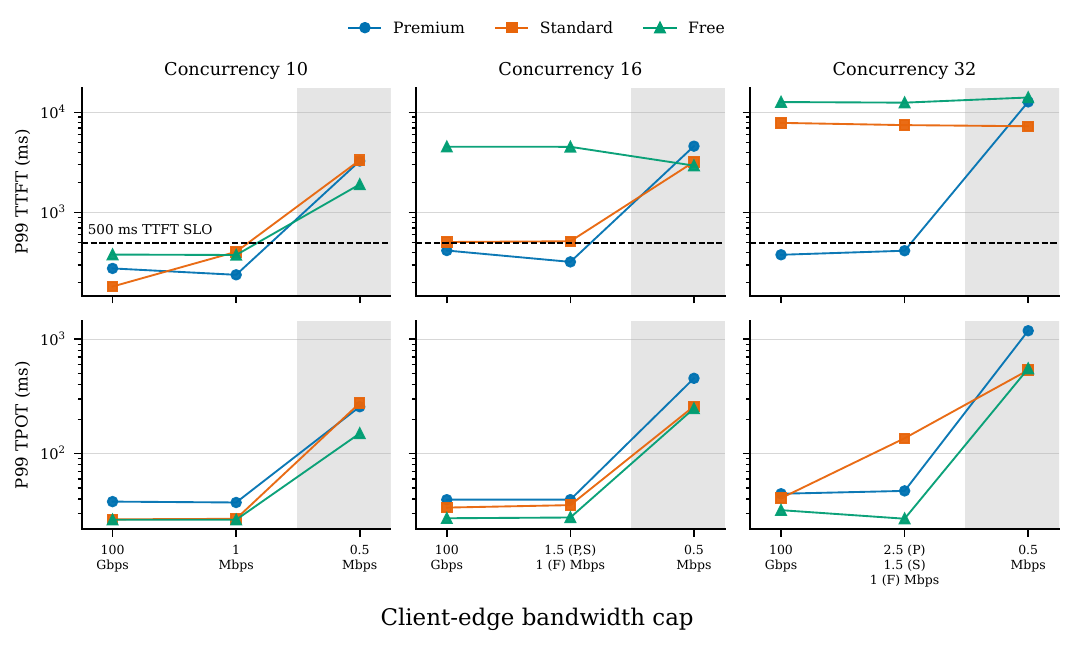}
\caption{Experiment N1: Effect of client-edge bandwidth caps on P99 \TTFT (top) and P99 \TPOT (bottom) at client concurrency 10, 16, and 32. The dashed line marks the 500~ms \TTFT \SLO, and the shaded region marks the 0.5~Mbps cap. The middle cap is set near each tier's unconstrained demand and differs by tier at concurrency 16 and 32. Full results are in Appendix~\ref{app:bw_sweep}.}
\label{fig:bw_discriminator}
\end{figure*}

\paragraph{Eval Experiment N1: End client bandwidth limiting}
Figure~\ref{fig:bw_discriminator} summarises the sweep across the three client concurrency levels, and the full per-tier results are given in Tables~\ref{tab:bw_sweep_c10} to~\ref{tab:bw_sweep_c32} in Appendix~\ref{app:bw_sweep}. At moderate load (concurrency 10), all three tiers are largely insensitive to bandwidth while the available capacity is at or above approximately 1~Mbps.
Reducing the cap to 1~Mbps leaves throughput (369--380~tok/s) and P99 \TPOT (26--37~ms) essentially unchanged, and P99 \TTFT remains below the 500~ms \SLO for every tier.
Reducing the cap to 0.5~Mbps produces a clear network bottleneck: the measured receive rate saturates at approximately 0.46~Mbps, throughput falls to 217--224~tok/s, P99 \TTFT rises to 1.9--3.3~s, and P99 \TPOT rises to 150--276~ms.
The transition is therefore bracketed between 0.5 and 1~Mbps at this load. Above it, inference and scheduling dominate latency; below it, network throughput becomes the binding resource and degrades both first-token delivery and subsequent token generation.

\begin{figure}
\centering
\includegraphics[width=0.9\columnwidth]{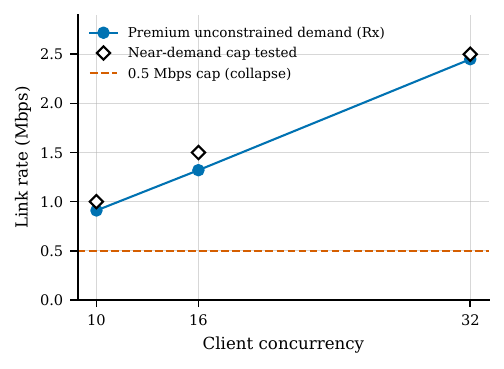}
\caption{Experiment N1: Premium unconstrained receive rate (measured at 100\,Gbps) against client concurrency, with the near-demand cap tested at each load and the fixed 0.5~Mbps cap. The bandwidth needed to remain in the unconstrained regime scales with offered load.}
\label{fig:bw_floor}
\end{figure}

The bandwidth required to stay in the unconstrained regime grows with load.
As shown in Figure~\ref{fig:bw_floor}, the unconstrained Premium receive rate rises from 0.91~Mbps at concurrency 10 to 1.32~Mbps at 16 and 2.45~Mbps at 32, and Premium stays within its \SLO under caps placed just above these values (1, 1.5, and 2.5~Mbps respectively).
At the tightest margin (concurrency 32, 2.5~Mbps cap, roughly 2\% above demand), \TTFT and \TPOT are preserved but throughput is about 7\% lower than the unconstrained baseline (934 versus 1001~tok/s), indicating that the link is close to becoming binding.
In contrast, the 0.5~Mbps cap is identical at every concurrency level, so it exposes the collapse but does not locate the floor: at all three loads the receive rate is pinned at 0.45--0.47~Mbps, throughput falls to 217--231~tok/s, and at concurrency 32 Premium P99 \TTFT reaches 12.8~s and P99 \TPOT 1.19~s (Figure~\ref{fig:bw_discriminator}).

Figure~\ref{fig:bw_discriminator} also separates network-limited from compute- or queue-limited behaviour.
At concurrency 16 and 32, Standard and Free already violate the \TTFT \SLO with an effectively unconstrained 100\,Gbps link (marginally for Standard at 16, by a wide margin otherwise), while their \TPOT remains close to the Premium baseline.
Their high \TTFT therefore cannot be attributed primarily to the network and instead reflects queueing or compute-side saturation, which the 0.5~Mbps cap then compounds.
Bandwidth effects can appear before the 0.5~Mbps floor when demand approaches the cap: Standard at concurrency 32 with a 1.5~Mbps cap (demand 1.44~Mbps) shows P99 \TPOT rising from 41 to 136~ms and its transmit rate roughly halving, even though its \TTFT is already dominated by queueing.
Taken together, these results distinguish the network bottleneck from the previously observed queue boundary: queue saturation predominantly increases \TTFT, whereas severe bandwidth saturation reduces measured link and token throughput and increases both \TTFT and \TPOT by an order of magnitude.

\paragraph{Eval Experiment N2: Distributed model latency characterisation (TP/PP across \PoP nodes)}

\begin{table*}
\centering
\caption{Experiment N2: GPT-OSS-20B serving latency and inter-shard
communication vs.\ inter-node RTT at each configuration's maximum
queue-free offered load (the Rate column; raised until median \TTFT
departs from the per-request baseline). Goodput = requests/s meeting
\TTFT$<$2\,s. Per-shard TX and
RX are Prometheus byte-counter rates in Mbps over the bench window,
mean (peak): for the \PP rows they are the sending and receiving
shard, for the \TP rows the per-shard values. All windows are
post-warm-up steady state.}
\label{tab:exp2_latency}
\footnotesize
\setlength{\tabcolsep}{3pt}
\begin{tabular}{llrrrrrrll}
\toprule
\textbf{Config} & \textbf{Pair} & \textbf{RTT} & \textbf{Rate} &
\textbf{Med.\ \TTFT} & \textbf{P99 \TTFT} & \textbf{tokens/s} &
\textbf{Goodput} & \textbf{Direction} &
\textbf{Per-shard TX / RX} \\
\midrule
\texttt{tp1} (intra) & single node & 0\,ms & 4.40 & 85\,ms & 326\,ms & 697 & 4.40 & local only & 1.0 (2.6) / 0.4 (1.6) \\
\texttt{pp2} & \augusta--\texttt{VM1@Soton} & 9\,ms & 2.60 & 150\,ms & 221\,ms & 403 & 2.60 & one-way & 28.7 (98) / 29.5 (54) \\
\texttt{pp2} & \augusta--\equestris & 18\,ms & 1.55 & 221\,ms & 286\,ms & 239 & 1.55 & one-way & 14.8 (39) / 13.4 (24) \\
\texttt{pp2} & \equestris--\claudiaxi & 33\,ms & 0.85 & 364\,ms & 466\,ms & 135 & 0.85 & one-way & 19.0 (44) / 17.6 (25) \\
\texttt{tp2} & \augusta--\texttt{VM1@Soton} & 9\,ms & 0.085 & 1.67\,s & 1.98\,s & 5.5 & 0.02 & all-reduce & 30 (54) / 29 (54) \\
\texttt{tp2} & \augusta--\equestris & 18\,ms & 0.09 & 2.9\,s & 3.6\,s & 3.1 & 0.00 & all-reduce & 18 (98) / 18 (99) \\
\texttt{tp2} & \equestris--\claudiaxi & 33\,ms & 0.035 & 4.8\,s & 5.7\,s & 1.8 & 0.00 & all-reduce & 11 (19) / 11 (19) \\
\bottomrule
\end{tabular}
\end{table*}

Table~\ref{tab:exp2_latency} reports end-to-end serving latency when
model shards span \PoP nodes, each row measured at its maximum
queue-free offered load. At this load \PP meets the 500\,ms \SLO on
every measured path: median \TTFT rises monotonically with link \RTT
(85\,ms co-located, then 150, 221 and 364\,ms at 9, 18 and 33\,ms) and
P99 stays below 0.5\,s. However, goodput degrades with distance. The
maximum queue-free rate falls from 4.40\,requests/s co-located to
2.60, 1.55 and 0.85\,requests/s, which is 59\%, 35\% and 19\% of the
co-located rate, and output throughput falls from 697 to 403, 239 and
135\,tokens/s. \PP therefore satisfies the latency \SLO but sustains
progressively less load as \RTT grows, so its feasibility at a given
distance is determined by the capacity it retains rather than by
latency alone. \TP collapses even at the lowest loads tested. Its
maximum queue-free rate is 0.035--0.09\,requests/s and its median
\TTFT of 1.7--4.8\,s exceeds the 500\,ms \SLO on every pair. At the
relaxed 2\,s threshold, goodput is zero on the 18 and 33\,ms pairs,
and the 9\,ms pair is the boundary case. The per-layer all-reduce on
the critical path is consistent with this cost. EP is likewise
infeasible: the \texttt{enable\_expert\_parallel} arms serve, but every
request times out on expert all-to-all. Four-shard arms (\texttt{pp4},
\texttt{tp2\_pp2}, \texttt{tp4}, \texttt{tp4\_ep4}) fail to initialise
within a 40-minute budget.

\paragraph{Eval Experiment N3: Distributed model bandwidth limiting}

\begin{figure*}
    \centering
    \includegraphics[width=\linewidth]{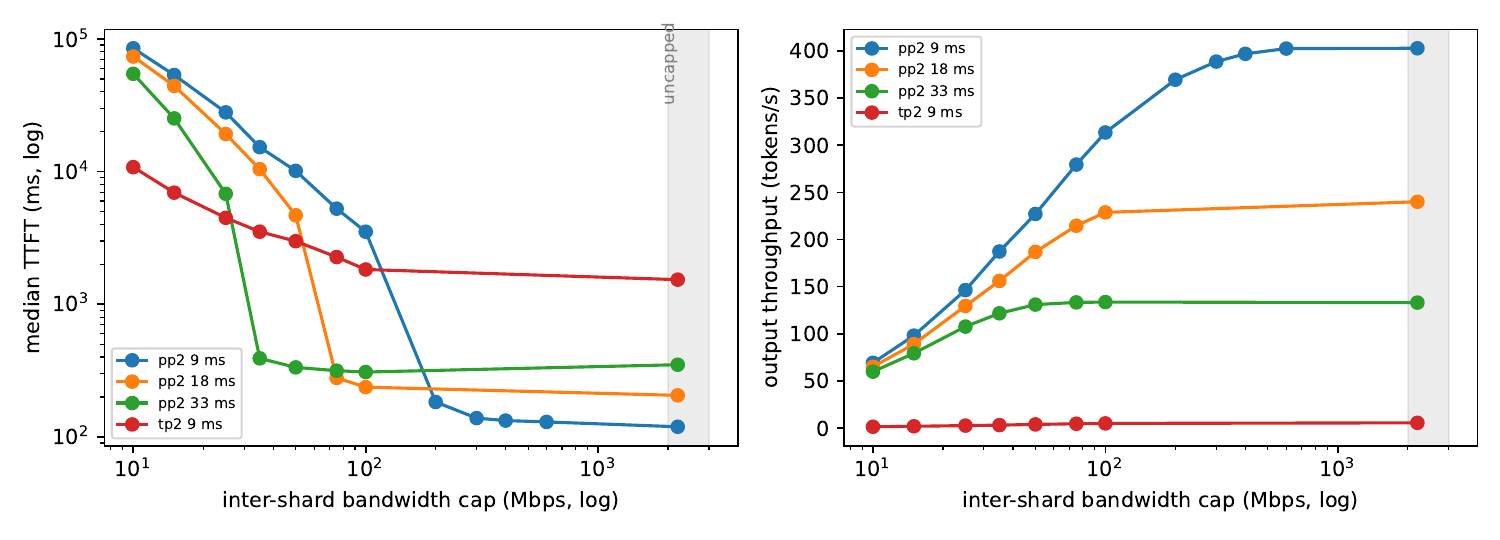}
    \caption{Experiment N3: median \TTFT and output throughput vs.\
             inter-shard bandwidth cap for \texttt{pp2} on the
             \augusta--\texttt{VM1@Soton}, \augusta--\equestris and
             \equestris--\claudiaxi pairs and for \texttt{tp2} on
             \augusta--\texttt{VM1@Soton}, GPT-OSS-20B. Each configuration
             is swept at its maximum queue-free offered load
             (Table~\ref{tab:exp2_latency}); the shaded region marks the
             effectively unconstrained setting.}
    \label{fig:exp3_bandwidth}
\end{figure*}

Figure~\ref{fig:exp3_bandwidth} shows that inter-shard bandwidth
limits \PP well before average traffic demand reaches the configured
cap. Each configuration is swept at its maximum queue-free offered load
(Table~\ref{tab:exp2_latency}). At fixed \RTT, median \TTFT remains
largely stable above a pair-specific threshold, then collapses abruptly
as the cap is reduced further, accompanied by a drop in throughput. The
threshold, defined as the cap at which median \TTFT doubles relative to
the unconstrained reference, lies between 100 and 200\,Mbps for the
9\,ms pair, 50 and 75\,Mbps for the 18\,ms pair, and 25 and 35\,Mbps
for the 33\,ms pair. In contrast, \TP exhibits no comparable cliff. Its
\TTFT remains broadly stable down to around 100\,Mbps and then rises
gradually, consistent with latency being dominated by the per-layer
all-reduce chain. These measurements are reliable, since unconstrained
references taken before and after every sweep agree and the 1\,Gbps
null control is indistinguishable from the unconstrained setting.

The \PP thresholds exceed the measured mean one-way traffic demand of
29, 15 and 19\,Mbps on the 9, 18 and 33\,ms pairs
(Table~\ref{tab:exp2_latency}). This gap reflects the bursty nature of
pipeline activations, since performance depends not only on average
bandwidth demand but also on whether each burst can drain before the
next arrives. As the cap is reduced, burst-drain time consumes more of
the inter-arrival interval, increasing \TTFT despite low average link
utilisation. The bandwidth floor therefore captures a burst-level constraint that mean traffic demand alone does not reveal.

\paragraph{Eval Experiment N4: Latency characterisation in Functional Disaggregation (P/D-D across \PoP nodes).}
\label{sec:evaluation:pdd}Table~\ref{tab:exp4_pd} reports serving latency for the disaggregated
deployments relative to the aggregated co-located baseline of the same
serving stack. All placements are driven at 2.7\,requests/s, the
maximum queue-free rate of the co-located baseline. Splitting prefill
and decode across \PoP nodes multiplies median \TTFT from 0.11\,s
co-located to 4.2--4.4\,s in every disaggregated arm, with P99 \TTFT at
approximately 19\,s, and output throughput falls by roughly a quarter.
The penalty is nearly invariant across pairs, since median \TTFT stays
within a narrow 4.2--4.4\,s band from 9 to 21\,ms of \RTT. This
indicates that the \KV-cache transfer and the NIXL handover dominate the
cost, while propagation delay plays little role. Median \TPOT is
essentially unchanged, so the queue-free per-token gain from
disaggregation is negligible.

The per-pod network counters on the \KV-sending pod show that the
transfer is not bandwidth-limited. Over the bench window the
disaggregated pair transfers 1.5--3.6\,GB of \KV-cache, or 15--35\,MB
per request. The stream is strictly one-way from prefill to decode,
with the return path carrying only a few percent of the volume. It
sustains 205--315\,Mbps on average (peaks of 481--855\,Mbps), which is
roughly 10--25\% of the 1.3--2.8\,Gbps these paths can carry. The cost
is therefore set by the handover itself and not by link capacity, and
because it does not grow with \RTT it is more consistent with a fixed
per-request overhead of the \KV connector than with a chain of
\RTT-bound transfer descriptors. Batching or pipelining the transfer
would be needed to reduce this delay.

\begin{table*}
\centering
\caption{Experiment N4: prefill/decode disaggregation across \PoP pairs
         versus the aggregated co-located baseline (Llama-3.1-8B-Instruct,
         ShareGPT, 100 prompts). All placements operate at an offered load of
         2.7\,requests/s, corresponding to the maximum queue-free rate.
         All deployments use the production serving stack (vLLM 0.12).
         \KV moved is the byte-counter volume on the \KV-sending pod over
         the bench window.}
\label{tab:exp4_pd}
\footnotesize
\setlength{\tabcolsep}{3pt}
\begin{tabular}{lrrrrrr}
\toprule
\textbf{Placement (pair \RTT)} &
\textbf{Med.\ \TTFT (s)} &
\textbf{P99 \TTFT (s)} &
\textbf{\TPOT (ms)} &
\textbf{tokens/s} &
\textbf{\KV moved (GB)} &
\textbf{peak/mean (Mbps)} \\
\midrule
co-located (0\,ms)              & 0.11 & 0.17 & 25.2 & 406 & --   & -- \\
\augusta--\texttt{VM1@Soton} (9\,ms)    & 4.24 & 19.0 & 24.3 & 297 & 1.54 & 567 / 205 \\
\augusta--\equestris (18\,ms)   & 4.37 & 19.1 & 25.1 & 308 & 3.55 & 855 / 315 \\
\texttt{VM1@Soton}--\equestris (21\,ms) & 4.44 & 19.6 & 25.2 & 297 & 1.95 & 481 / 208 \\
\bottomrule
\end{tabular}
\end{table*}



\section{Discussion and Open Questions}
\label{sec:discussion}

\subsection{Key Lessons}
\subsubsection{L1: Communication Patterns Define the Operating Boundaries of Distributed Inference}

Table~\ref{tab:distributed_feasibility} summarises the resulting
operating boundaries: which distributed inference techniques remain
feasible as compute is spread across increasingly distant sites. We
deem a technique feasible at a tier if it meets the 500\,ms P99 \TTFT
\SLO at a queue-free offered load that retains at least 50\% of the
co-located maximum queue-free rate. The first condition captures
latency compliance and the second captures sustainable capacity. The
reach of each technique is determined by its communication pattern
rather than by available bandwidth alone. Tensor parallelism, which
places an all-reduce on the critical path at every layer, does not
remain feasible beyond a single subnet. Pipeline parallelism and
prefill--decode disaggregation, which exchange activations and
\KV-cache state respectively, avoid synchronous all-reduce semantics
and therefore tolerate wider-area placement in principle. Within our
measurements, pipeline parallelism meets the latency \SLO on every
measured cross-\PoP pair (median \TTFT 85--364\,ms, P99 $<$0.5\,s,
Table~\ref{tab:exp2_latency}), but the load it can sustain falls with
\RTT: the maximum queue-free rate is 2.60, 1.55 and 0.85\,requests/s
at 9, 18 and 33\,ms, which is 59\%, 35\% and 19\% of the co-located
4.40\,requests/s. Under the criterion above, \PP is therefore feasible
at the Mid tier but not at the Far and Distant tiers, where it
satisfies the latency \SLO only at a fraction of the co-located
capacity. The picture is also load-regime dependent: under saturation,
client-side queueing inflates \PP \TTFT to 7.2--23.1\,s on the same
pairs. Configurations spanning more than one hop could not be
initialised within our deployment budget. Prefill--decode
disaggregation does not meet the \SLO at any cross-\PoP tier: its
penalty is set by the \KV-cache transfer and handover rather than by
propagation delay, and median \TTFT stays within 4.2--4.4\,s across
9--21\,ms. Data parallelism, which introduces no inter-node
communication on the critical path, is insensitive to placement, but
requires the model to fit within a single node's GPU memory and
therefore does not generalise to models exceeding that bound.

\subsubsection{L2: Bottlenecks Shift with Load}

The resource that limits SLO attainment is not fixed: it changes with
offered load, and the three regimes we observed leave distinct
signatures in the latency metrics. At low load, the three tiers are
indistinguishable and inference and scheduling dominate latency
(Section~\ref{sec:evaluation}). As load grows, each slice reaches the
capacity set by its replica ceiling and becomes queue-limited: \TTFT
rises by an order of magnitude while \TPOT remains close to its
unconstrained value (Tables~\ref{tab:bw_sweep_c10} to~\ref{tab:bw_sweep_c32}). When
the client-edge bandwidth cap is tightened, the slice becomes
network-limited instead: measured link and token throughput clamp, and
both \TTFT and \TPOT increase by an order of magnitude. The bandwidth
needed to stay in the unconstrained regime also grows with load (0.91,
1.32 and 2.45\,Mbps at concurrency 10, 16 and 32 for the Premium
tier), so a cap that is sufficient at one load becomes the binding
constraint at the next. A similar pattern appears across \PoP nodes:
\PP meets the latency \SLO at bounded load, but under saturation
client-side queueing, not \RTT, inflates its \TTFT to 7.2--23.1\,s.

The practical implication is that no single utilisation metric is sufficient to identify the resource constraining SLO attainment. Instead, the control plane (Section~\ref{sec:design:control}) must distinguish queue saturation from network saturation by jointly interpreting \TTFT, \TPOT and link throughput, supported by the cross-layer observability described in Section~\ref{sec:design:observability}. This distinction is essential for effective resource allocation: adding replicas can alleviate queueing but cannot resolve a bandwidth bottleneck. Resource scaling must therefore be guided by the identified bottleneck, rather than triggered by utilisation alone.

\begin{table*}
\centering
\caption{Feasibility of distributed inference techniques across JOINER
         cross-\PoP latency tiers. A technique is feasible if it meets
         the 500\,ms P99 \TTFT \SLO at a queue-free offered load that
         retains at least 50\% of the co-located maximum queue-free
         rate. TP and PP evaluated with GPT-OSS-20B; DP and P/D-D
         evaluated with Llama-3.1-8B-Instruct. All deployments
         constrained by L40S GPU memory. \cmark: feasible,
         \xmark: unfeasible.}
\label{tab:distributed_feasibility}
\begin{tabular}{lcc cccc}
\toprule
\multirow{2}{*}{\textbf{PoP tier}} &
\multirow{2}{*}{\shortstack{\textbf{RTT}\\\textbf{(ms)}}} &
\multirow{2}{*}{\shortstack{\textbf{TCP TX}\\\textbf{(Gbps)}}} &
\multicolumn{4}{c}{\textbf{Distributed AI technique}} \\
\cmidrule(lr){4-7}
 & & & \textbf{TP} & \textbf{PP} & \textbf{DP} & \textbf{P/D-D} \\
\midrule
Intra-PoP (same subnet)          & 0.60  & 10.0 & \cmark & \cmark & \cmark & \cmark \\
Near cross-PoP ($<$5\,ms)        & 4.26  & 6.53 & \xmark & \cmark & \cmark & \xmark \\
Mid cross-PoP (5--10\,ms)        & 7.57  & 2.79 & \xmark & \cmark & \cmark & \xmark \\
Far cross-PoP (10--15\,ms)       & 10.90 & 1.19 & \xmark & \xmark & \cmark & \xmark \\
Distant cross-PoP ($\geq$15\,ms) & 18.97 & 1.30 & \xmark & \xmark & \cmark & \xmark \\
\bottomrule
\end{tabular}

\vspace{2pt}
\footnotesize
\textbf{TP}: shards layers across nodes with per-layer all-reduce on the
critical path; lowest memory per node, highest latency sensitivity.\\
\textbf{PP}: partitions by layer group, exchanging only activations;
meets the latency \SLO at every measured \RTT, but its sustainable
queue-free rate falls to 59\%, 35\% and 19\% of the co-located rate at
9, 18 and 33\,ms.\\
\textbf{DP}: replicates the model with no inter-node traffic;
placement-insensitive, but the model must fit in a single node's GPU
memory.\\
\textbf{P/D-D}: separates prefill and decode, transferring the
\KV-cache between them; unfeasible at every cross-\PoP tier, with an
\RTT-insensitive penalty set by that transfer.
\end{table*}

\subsubsection{L3: Cold Start Limits Elastic SLO Enforcement}

Elastic scale-out only protects an \SLO if new replicas become ready
before the queue builds. In our deployment, weight acquisition rather
than scheduling bounds this delay: pods reach readiness within about two
minutes under both loaders, so the remaining time is spent loading
weights and producing the first token. Even with parallel weight
streaming, which reduces end-to-end scale-out from approximately
1{,}500\,s to 610\,s, a replica added under load takes minutes to serve
its first response, whereas the \SLO breaches we measured develop within
a single step of the concurrency sweep. The transient \TTFT excursion
of the Premium tier at concurrency 16 (Section~\ref{sec:evaluation})
shows the same effect at small scale. Hot-phase scale-out matches the
cold phase, so the delay is set by the loader and not by cache warmth,
and keeping weights cached on the node does not remove it.

Reactive autoscaling alone therefore cannot enforce latency \SLO{}s
under load growth, even when it is configured with generous replica
ceilings. Capacity has to be provisioned ahead of demand, through
traffic forecasting, warm standby replicas, or faster weight
distribution, and the cold-start time of the largest model a tier
serves sets the lead time that forecasting must achieve. Our
measurements use an 8B model; the lead time will grow with model size,
which makes this constraint more severe for the models that most need
distributed serving.

\subsection{Scope and Limitations}

Our experiments use independent, short-lived, stateless chat requests drawn from ShareGPT, served on a homogeneous pool of NVIDIA L40S GPUs. Compute heterogeneity is emulated through Run:ai's fractional GPU scheduler. We use Llama-3.1-8B-Instruct for service differentiation experiments and GPT-OSS-20B for distributed-inference experiments. The operating boundaries reported in Table~\ref{tab:distributed_feasibility} therefore characterise a specific workload class on a single accelerator generation, under the evaluated resource configurations. Agentic workflows introduce chained, stateful calls with growing contexts, shifting the optimisation objective towards task-level latency and making \KV-cache locality an additional placement constraint. Longer-context workloads may consequently alter these operating boundaries, particularly for prefill/decode disaggregation, where \KV-cache transfer costs increase with the size of the context to be transferred. These scenarios, together with closed-loop control-plane adaptation to workload dynamics, remain outside the scope of the present evaluation. 

Service tiers were realised through statically configured per-slice policies, so the evaluation shows that coordinated policy enforcement yields distinct and predictable regimes, but not that the control plane can select configuration and serving location dynamically under changing load; the planning, forecasting and routing components of the architecture (Figure~\ref{fig:architecture}) were not evaluated in closed loop. Each distributed-inference technique was likewise evaluated in isolation, with network and compute dimensions varied one at a time (for example, rate-shaping inter-shard bandwidth at a fixed \RTT), so interactions between techniques and between simultaneous network and compute pressure are not captured. Finally, model-parallel results cover at most two shards, since four-shard configurations failed to initialise within our time budget. These limitations motivate the open questions that follow.

\subsection{Open Questions}

\subsubsection{Heterogeneity in compute and network}
Operator infrastructure is inherently heterogeneous: sites differ in accelerator generation, memory capacity and available headroom, while WAN paths span more than an order of magnitude in latency and vary nearly as much in throughput (Table~\ref{tab:network_profile}). Our results show that serving location is a first-order determinant of SLO attainment, making placement across heterogeneous sites a problem that existing serving systems have yet to address. We encourage the community to explore placement and routing policies that treat compute capability and network conditions jointly: when a nearby but less capable site should be preferred over a distant, better-provisioned one, how inference configurations and replica counts should be matched to mixed accelerators, and how heterogeneous replicas should interact with SLO-aware routing and autoscaling. Benchmarks that reflect heterogeneous, operator-controlled infrastructure rather than uniform clusters would make this research comparable and reproducible.

\subsubsection{Composability}
Our results establish distinct operating boundaries for distributed inference techniques over a WAN (Table~\ref{tab:distributed_feasibility}): replication is insensitive to placement, pipeline parallelism is feasible up to mid-tier latencies, and tensor parallelism is confined to a subnet. These boundaries suggest a hierarchy that matches each technique to the network tier it tolerates: tensor parallelism within a node, pipeline parallelism across nearby nodes, replication across sites, and prefill/decode disaggregation within each replica (Section~\ref{sec:background}). However, the feasibility of individual techniques does not establish the feasibility of their composition. How far such compositions extend the feasible region, and how they interact, is a rich space for the community to explore. For example, replica-level autoscaling may amplify pipeline bubbles, while \KV-cache transfer layered onto pipelined stages may introduce additional communication dependencies. A control plane that selects compositions according to site capabilities, network conditions and \SLO{}s, accounting for the interactions between techniques rather than optimising each independently, is a natural next step.

\subsubsection{Serving models that exceed single-node memory over the WAN}
Our measurements identify communication on the critical path as the principal barrier to scaling out over the WAN. Replication is insensitive to placement but requires the model to fit on each serving node. Tensor parallelism is confined to a single subnet by per-layer all-reduce, while pipeline parallelism meets the latency \SLO across sites but sustains less load as \RTT increases, and is limited by the burst latency of activation transfers rather than by average bandwidth. Prefill/decode disaggregation encounters a different bottleneck: transferring 15--35\,MB of \KV-cache per request adds seconds to \TTFT even though the transfer occupies only a small fraction of the available bandwidth, indicating that message-level latency, rather than throughput, constrains the handover. These findings point to concrete directions for inference systems research: overlapping communication with computation, compressing and shaping activation bursts, batching and pipelining \KV-cache transfer descriptors, and relaxing synchrony through asynchronous or speculative execution. Progress here would allow operators to serve models that exceed the memory of any single site, which is currently out of reach over a WAN.

\subsubsection{Stateful and agentic workloads}
Our experiments use stateless chat requests, whereas agentic applications chain stateful calls with growing contexts, shifting the service objective from per-request latency to task-level latency. In this setting, \KV-cache locality becomes a placement constraint: routing a follow-up call to a site that already holds the relevant cache may be preferable to selecting the site with the most spare capacity. Our disaggregation results show that moving 15--35\,MB of \KV-cache per request over the WAN adds seconds to \TTFT (Section~\ref{sec:evaluation:pdd}), so placement must account for cache residency and transfer cost as well as network conditions. Open questions include how \SLO{}s should capture task completion time and tail latency across multi-step workflows, how routing should balance cache affinity against load and failure recovery, and how distributed \KV-cache management should support reuse, migration and eviction across sites as contexts grow. Exploring evaluation methodologies and serving policies designed for stateful, agentic workloads over geographically distributed, operator-controlled infrastructure are key aspects for broader impact and adoption.

\subsubsection{SLO-aware placement and control}
The control plane of InferSlice includes planning and optimisation, traffic forecasting, semantic routing, prompt caching and distributed \KV-cache components (Figure~\ref{fig:architecture}), whereas our evaluation exercises statically configured per-slice policies. Our results show that admission limits, replica ceilings and serving location each shape \SLO attainment, and that the binding resource changes with load. The central open problem is closing the loop between observation, prediction and resource control, which requires forecasting demand accurately enough to scale ahead of cold-start delays, identifying the binding resource online, and jointly selecting serving locations, inference configurations and replica counts as compute and network conditions evolve, while preserving differentiated \SLO{}s under load without over-provisioning. Energy, carbon and power availability add further objectives and constraints to this decision, as discussed next. We encourage the community to explore closed-loop, cross-layer control that integrates these decisions, since it is the point at which the other open questions come together.

\subsubsection{Energy-aware AI grid}
The geographic distribution of inference infrastructure also raises fundamental questions about the energy efficiency of AI serving. Rather than concentrating compute in purpose-built, gigawatt-scale AI factories, operators could aggregate existing capacity across data centres, edge sites and other infrastructure into a geographically distributed AI grid, improving the utilisation of underused resources while placing workloads where compute capacity, network conditions and energy availability are favourable. However, the energy implications cannot be inferred from accelerator utilisation alone, since distributed inference introduces network energy costs, replication increases the number of active accelerators, and the communication and synchronisation overheads characterised in this work can offset the gains from better utilisation. The same \SLO can therefore be met at very different energy cost depending on the inference configuration and placement chosen, and powering down idle replicas must be weighed against the cold-start lead times that forecasting has to achieve. This motivates extending \SLO-aware placement to jointly optimise compute, network and energy, including energy-aware routing, workload migration and consolidation, and the trade-off between local utilisation and \WAN communication. 

Carbon-aware orchestration adds a further dimension, because sites draw on grids with different and time-varying carbon intensity, so flexible, throughput-oriented tiers could be deferred or shifted towards cleaner sites and periods, within the limits set by sovereignty, latency and the heterogeneity of sites discussed above. Finally, power supply must be treated as a first-class resource alongside compute and network capacity. Telco sites were provisioned for network equipment rather than dense accelerators, so available power, cooling and backup capacity may bound the GPU capacity each site can host and sustain, and sites with limited headroom may be less suitable than better-provisioned but more distant ones. Exploiting this requires power and energy telemetry to be exposed through the observability layer, so that the control plane can reason over latency, energy, carbon and power availability when selecting serving locations, configurations and replica counts. We encourage the community to explore such energy- and carbon-aware orchestration, and to report energy alongside \SLO attainment in future evaluations.

\section{Conclusion}%
\label{sec:conclusion}

This paper presented InferSlice, a reference architecture for SLO-driven LLM inference over a telco WAN. InferSlice integrates network slicing, GPU partitioning, elastic serving and end-to-end observability, extending service differentiation from network connectivity to AI inference itself. Deployed across six PoPs of the JOINER testbed, on a shared network not dedicated to AI and without RDMA or specialised interconnects between sites, it shows that operator-controlled infrastructure can meet sovereignty, latency, and resource-locality requirements. Service tiers yield distinct latency-throughput regimes over the same GPU pool, with peak SLO-compliant goodput of 2.2$\times$ between the Premium and Free tiers, and parallel weight streaming reduces end-to-end scale-out time from approximately 1{,}500\,s to 610\,s. Our experiments also show that the binding resource shifts with load, with queue saturation and bandwidth saturation leaving distinct signatures in \TTFT and \TPOT, so that cross-layer observability is a prerequisite for SLO enforcement.

Over the WAN, the reach of each distributed inference technique is set by its communication pattern rather than by bandwidth alone. Replication is insensitive to placement but bounded by single-node memory, pipeline parallelism meets the 500\,ms SLO on every measured cross-site link while sustaining progressively less load as \RTT grows, and tensor parallelism and prefill/decode disaggregation are not feasible beyond a single subnet. In the case of disaggregation, the penalty is set by the \KV-cache handover rather than by propagation delay or available bandwidth. These boundaries suggest that distributed and heterogeneous inference systems should match each technique to the network tier it tolerates, and that serving models beyond the memory of a single site over a WAN remains an open problem. Our evaluation used statically configured policies, stateless chat workloads and a single accelerator generation, so closed-loop control, stateful and agentic workloads, heterogeneous accelerators and energy-aware placement are part of our future direction. We hope that the lessons and open questions presented here encourage further research on SLO-driven inference over geographically distributed, operator-controlled infrastructure.

\ifdefined\isaccepted
\section*{Acknowledgements}
The authors thank the NVIDIA Run:ai team for their technical support throughout this work, including assistance with platform deployment, early access, and the provision of reference examples that informed the design presented in this paper.
This work was partially supported by
the UK's \EPSRC~-- JOINER project, EP/Y036514/1 -- and
the Smart Networks and Services Joint Undertaking -- ECO-eNET project, Grant Agreement No. 101139133.

\fi
\section*{Use of Generative AI Declaration}
During the preparation of this work, the authors used Grammarly, Claude, and ChatGPT for grammar checking, and language and format editing. The authors subsequently reviewed and edited the content as necessary and take full responsibility for the content of this publication.

\bibliography{bibliography}
\bibliographystyle{mlsys2025}

\section*{Appendix}
\label{sec:appendix}

\appendix

\section{JOINER: Telco WAN testbed}
\label{sec:annex-joiner}

\begin{figure}
\centering
\includegraphics[trim={5pt} {8pt} {5pt} {18pt},clip,width=\columnwidth]{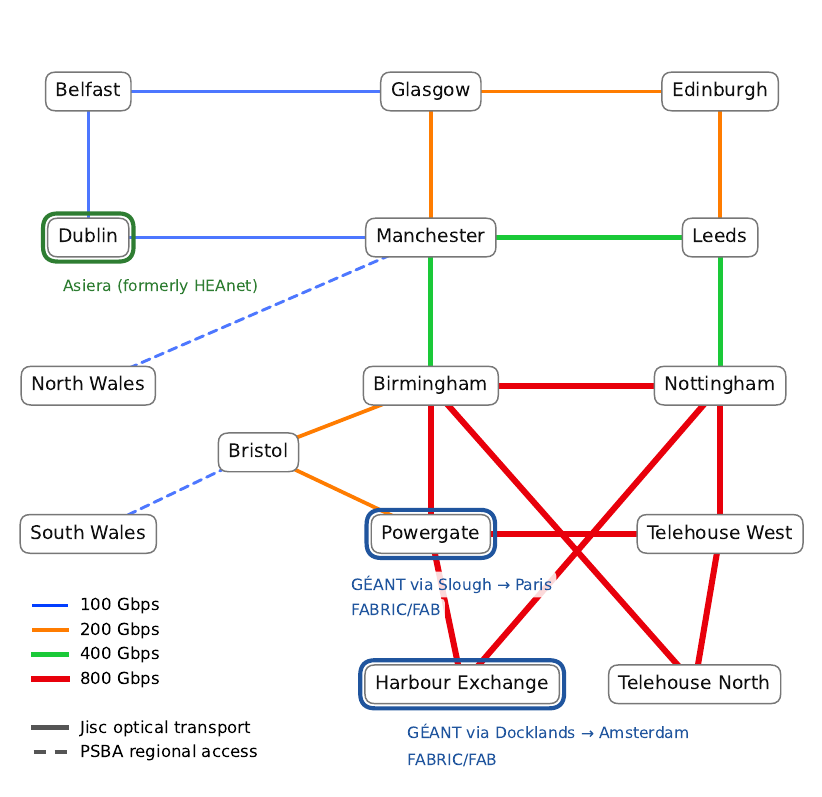}
\caption{The JANET backbone and its international research-network links (nominal link capacity 100 to 800~Gbit/s).}
\label{fig:infra-janet}
\end{figure}

The \acf{JOINER}
started as a research project in December 2023 with the goal of creating
a nationwide experimentation platform focused on telecom innovation and future connectivity.
Funded by the UK's \EPSRC as part of the Federated Telecoms Hubs,
it started public operations in 2025~\cite{joiner_launch} focusing on interconnecting and fostering
collaborations across participating institutions.
Initially bringing together 13 universities and research laboratories across the
UK, \JOINER announced its first international node in Ireland still in 2025~\cite{joiner_first_international}.
As of 2026, the list of institutions that are part of \JOINER is:

\begin{itemize}
  \setlength{\itemsep}{0pt}
  \setlength{\topsep}{0pt}
  \setlength{\parsep}{0pt}

  \item \TCD
  \item \QUB
  \item University of Glasgow
  \item University of Edinburgh
  \item Bangor University
  \item University of Leeds
  \item Cranfield University
  \item University of Cambridge
  \item \DC
  \item \UCL
  \item Imperial College London
  \item University of Oxford
  \item University of Southampton
  \item University of Bristol
\end{itemize}

Beyond the objective of federating research facilities,
\JOINER offers unique capabilities to the affiliated institutions
for accelerating research and development, including access to a hybrid cloud/edge with
collocated computing resources at each node deployed in a distributed fashion
and O-RAN compatible \RICs and packet cores.
In addition to these shared resources,
research in the JOINER project also culminated in the creation and
integration of the National Spectrum Facility, the AI Acceleration Facility and
the \NTN~Emulation Facility.

This combination of federated research capabilities, strong collaboration and
production-scale national and international breadth positions JOINER to support
the development of innovative technologies.
By using JOINER, partners are able to test prototypes in a realistic
large-scale environment, which helps to identify opportunities and issues that
otherwise would remain hidden until the technology is deployed as part of a
public-facing product. This is beneficial not only to the research community
but also industry, operators, vendors and policymakers.

\begin{figure*}
\centering
\includegraphics[width=0.9\linewidth]{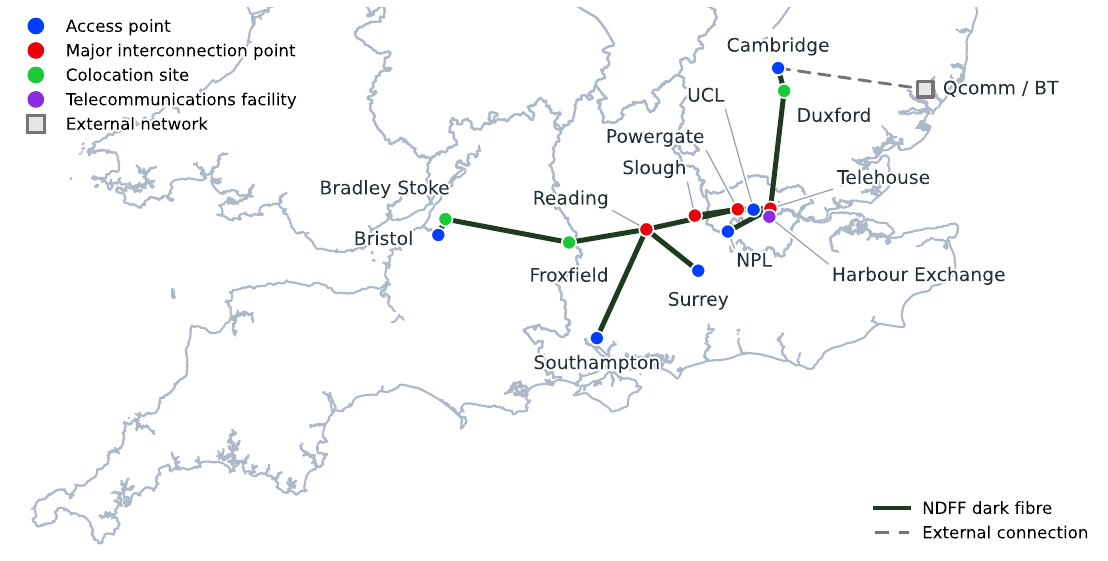}
\caption{%
  \acuse{NPL}\acuse{UCL}
  The \acl{NDFF} and its external links.
  Access points include \JOINER nodes in Bristol, Cambridge, Southampton and UCL,
  through which optical experiments can be carried out directly in the physical layer.
}
\label{fig:infra-ndff}
\end{figure*}

\subsection{Architecture}

\JOINER infrastructure is implemented in the form of a distributed testbed, where
each affiliated institution is assigned a terminal containing hardware to
support the experimental capabilities, which generally
comprises the following equipment~\cite{joiner_magazine}:

\begin{itemize}
  \setlength{\itemsep}{0pt}
  \setlength{\topsep}{0pt}
  \setlength{\parsep}{0pt}

  \item Oscilloquartz OSA 5412 for precision time distribution
    (\acs{PTP}/\acs{SyncE}) and \textbf{clock synchronisation}.
  \item high-performance Edgecore \acs{P4} switches based on the Intel Tofino
    chipset, complemented by AMD Alveo V80 accelerator cards for hardware
    offload and advanced \textbf{network programmability}.
  \item Juniper ACX routers and EX switches, together with Fortinet firewalls for
    \acs{SDWAN} deployment and \textbf{edge routing and securing}.
  \item Dell R760 and R660 servers with up to 64 physical cores, 512~GB+ RAM for
    \textbf{edge computing}, and Nvidia L40S GPUs for \textbf{AI acceleration}.
  \item Dell ME5024 \acs{SAN} arrays offering tiered \textbf{storage}
    (\acs{SSD}/\acs{HDD}) for high-throughput data access.
  \item Some sites can host additional \textbf{specialised hardware}, including
    \acs{RFSoC}~\acs{FPGA} evaluation kits, O-RAN compatible radios, and
    access to satellite links.
\end{itemize}

Control and management functionalities, such as remote access, central
authentication and authorisation and policy management, are aggregated via a
central node called \textit{Brain}, which also manages and automates the hybrid
cloud. In turn, the hybrid cloud is composed of VMware
\TCA\footnote{\url{https://www.vmware.com/products/cloud-infrastructure/telco-cloud-platform}}
(managing the on-premises compute resources at all nodes) paired
with complementary remote cloud services provided by \AWS\footnote{\url{https://aws.amazon.com/}},
Microsoft Azure\footnote{\url{https://azure.microsoft.com/}} and
\GCP\footnote{\url{https://cloud.google.com}}.
Additionally, \JOINER is also able to automate physical servers via
\MAAS\footnote{\url{https://canonical.com/maas}} and deploy open-source
virtualisation and container platforms, such as
OpenStack\footnote{\url{https://www.openstack.org}} and
OpenShift\footnote{\url{https://www.redhat.com/en/technologies/cloud-computing/openshift}}, on demand.

\subsection{Connectivity}

\JOINER terminals provide access to data-link (layer 2 in the \OSI model)
and physical connectivity via a series of national and international
partnerships with research and commercial networks.


At its core, \JOINER relies on JANET\footnote{\url{https://www.jisc.ac.uk/janet}},
the UK's \NREN, operated by Jisc, a not-for-profit organisation responsible
for managing and operating education and research digital infrastructure.
Figure~\ref{fig:infra-janet} shows JANET's major interconnection points~\cite{jisc_janet_update_2025}
and its access to other international research networks like the European
GÉANT\footnote{\url{https://network.geant.org}} and the USA-led
FAB\footnote{\url{https://fabric-testbed.org/about/fab}}, through which it is
also possible to connect to other institutions in Europe, North and South America and Asia.
Regional access in Wales is provided via the \PSBA, which links sites across North and South Wales
to JANET's major interconnection points in the southwest and northwest of England.
In Ireland, JANET also interconnects to the Irish \NREN administrated by Asiera
(formerly HEAnet).

Furthermore, \JOINER is also connected to the \NDFF which supplies direct
optical connectivity between some of the terminals (Bristol, Cambridge,
Southampton and UCL)~\cite{ndff_what_is} and enables optical experiments in both classical and
quantum communications.
Figure~\ref{fig:infra-ndff} details the physical layer, tracing the dark-fibre
corridor across the southern part of England, along with an external
link to BT's premises in Adastral Park.

\JOINER also supports \NTN connectivity with links carried
over \LEO and \GEO services, and includes a nomadic terminal (assembled in a
customised vehicle) that is backhauled via satellite and carries an on-board 5G
radio access and core, with up to five hours of off-grid operation~\cite{joiner_architecture}.

\begin{figure*}
\centering
\includegraphics[trim={40pt} {50pt} {45pt} {95pt},clip,width=0.7\linewidth]{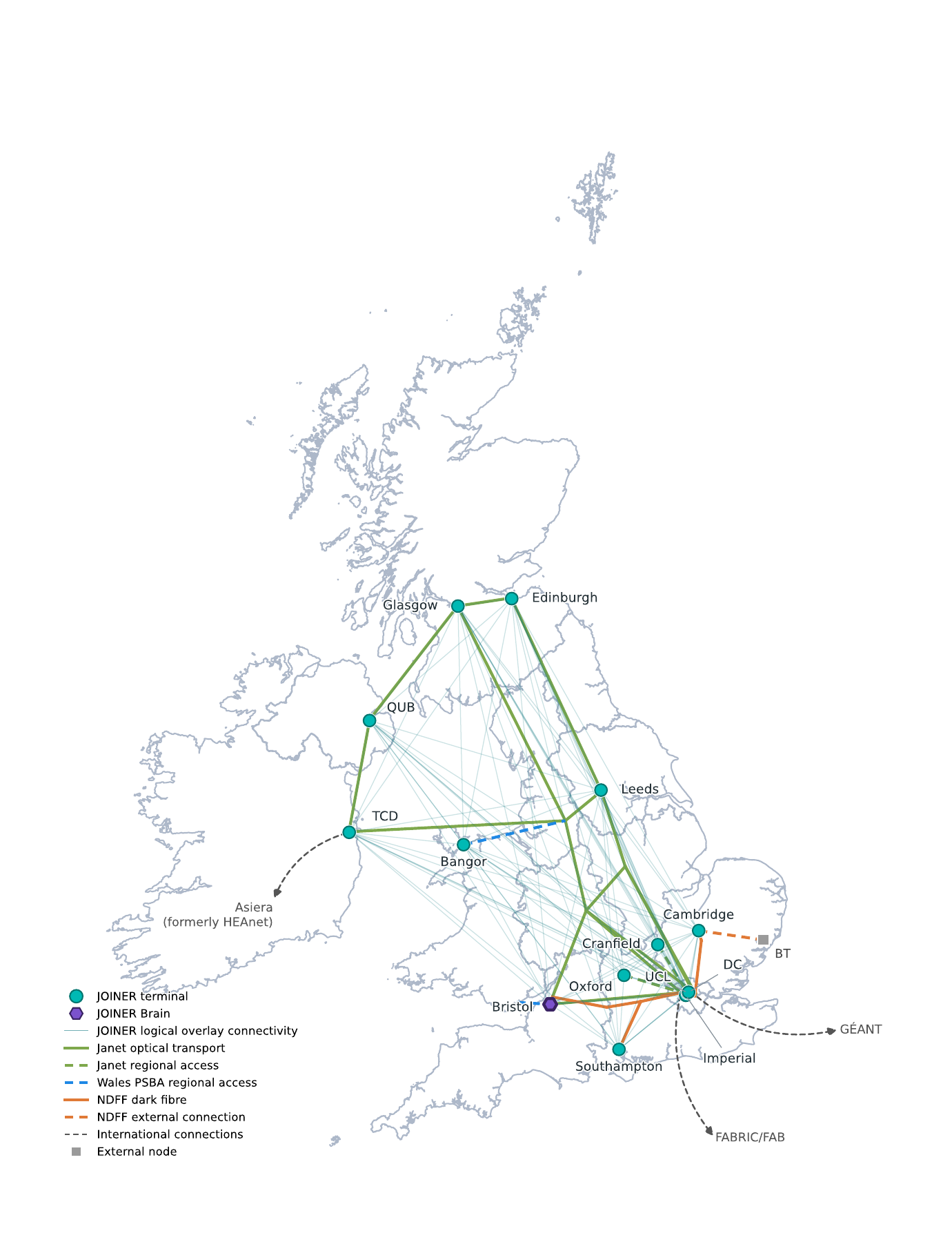}
\caption{
  \acuse{DC}\acuse{UCL}\acuse{TCD}\acuse{QUB}
  The 14 \JOINER sites as a federated overlay on top of JANET and \NDFF,
  with the JOINER Brain colocated at Bristol and international connections to
  GÉANT, FABRIC/FAB and Asiera.
}
\label{fig:testbed-fullmesh}
\end{figure*}

On top of the physical underlay, \JOINER provides multi-tenant network connectivity.
Terminals interconnect at~10--40~Gbit/s in a logical
mesh~\cite{joiner_architecture}, delivered as \SDWAN over \LtwoVPN.
Different research projects and experiments can run simultaneously, and each one is able to
deploy its own logical overlay and provision multiple isolated data-link layer
networks mapped into \VLANs over the same shared infrastructure.
Figure~\ref{fig:testbed-fullmesh} consolidates the JANET and \NDFF connectivity
into the \JOINER platform's complete federated view,
illustrating the overlay connectivity as a single mesh of dynamic \VLAN-isolated links.

\section{InferSlice dashboard description}
\label{sec:dashboard}

\begin{figure*}
    \centering
    \includegraphics[width=\linewidth]{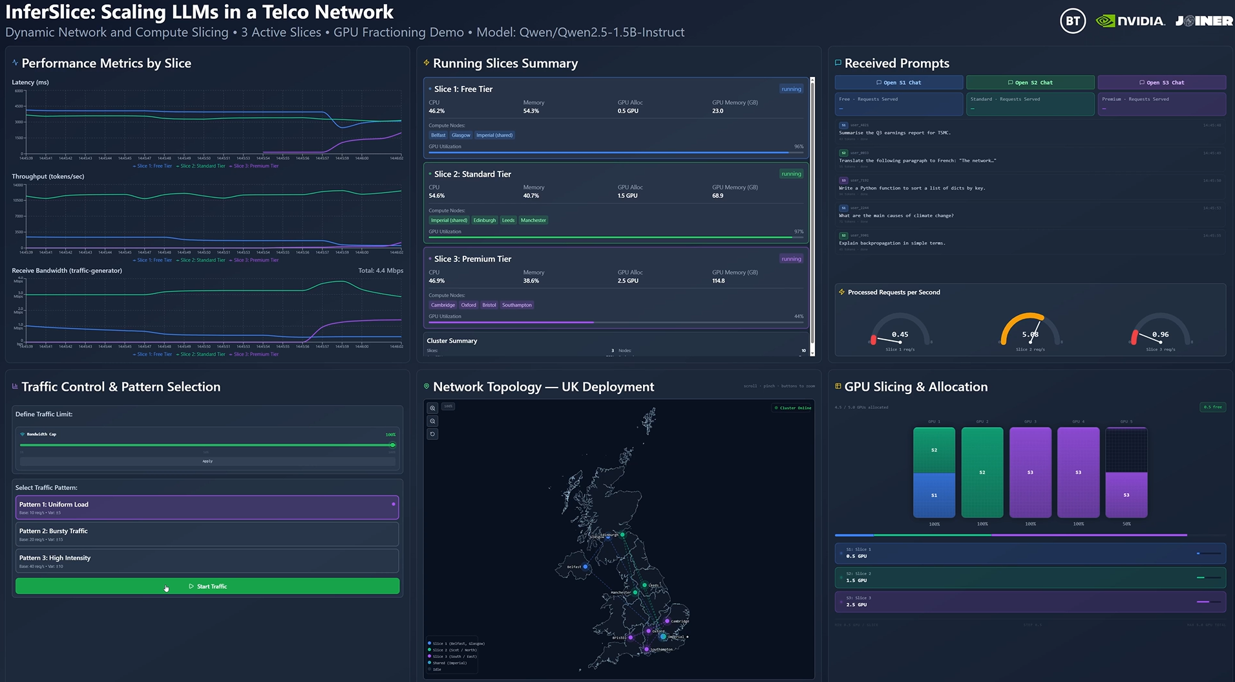}
    \caption{InferSlice dashboard for real-time user interaction and monitoring. The dashboard provides interactive traffic generation, performance monitoring, resource allocation visualisation, and network topology information across premium, standard, and free service tiers. The interface enables users to observe the impact of traffic patterns and bandwidth constraints on autoscaling behaviour and resource distribution in real time.}
    \label{fig:dashboard}
\end{figure*}

For demonstration purposes, a user-friendly dashboard was developed to enable real-time interaction with the InferSlice setup. Users can interact with the dashboard and send predefined traffic patterns to the LLM endpoints, allowing them to observe the system's real-time behaviour and performance. Autoscaling and resource allocation behaviour can be monitored according to the system's current load, alongside real-time metrics such as latency, throughput, and network bandwidth. The dashboard focuses on comparing information across the premium, standard, and free tiers. Figure~\ref{fig:dashboard} shows the dashboard. Experiments are conducted across three tiers, each corresponding to a different network slice. The premium, standard, and free tiers are represented by purple, green, and blue, respectively. 

The dashboard presents real-time information across six panes. \textbf{Top-left: Performance metrics by slice.} This pane displays graphs of real-time inference performance metrics, including latency (ms), throughput (tokens/s), and network bandwidth (Mbps), over a fixed time window. Metrics from all tiers are shown on the same graphs to facilitate comparison. \textbf{Bottom-left: Control and pattern selection.}  This pane allows users to interact with the dashboard and manually generate LLM traffic in real time. Users can specify the network bandwidth cap and select the traffic pattern to be generated. Three traffic patterns are available: Uniform Load, Bursty Traffic, and High Intensity. The behaviour of the InferSlice system is influenced by these user-defined parameters and is reflected throughout the dashboard. \textbf{Top-centre: Running slices summary.} This pane displays hardware resource utilisation by tier. The reported resources include CPU usage, memory usage, GPU allocation count, and GPU memory utilisation. The compute nodes associated with each tier are also displayed. \textbf{Bottom-centre: Network topology.} This pane presents the geographical locations of the compute nodes on a map of the UK, together with the network connections between them. Nodes and connections are colour-coded according to their respective tiers. \textbf{Top-right: Received prompts.} This pane displays a sample of the real-time prompts currently being processed by the LLM. Prompts are categorised according to their corresponding tiers. Speedometer-styled gauge charts display the request processing rate for each tier. \textbf{Bottom-right: GPU slicing and allocation.} This pane displays the occupancy of available GPUs in the cluster across different tiers. Fractional GPUs allocation is supported for each tier. Premium tier can occupy the largest share of GPU resource, followed by the standard tier and then the free tier.

\section{Extended experimental evaluation}
\subsection{Concurrency-policy differentiation}
\label{app:concurrency_policy}

Section~\ref{sec:evaluation:tiers} differentiates the slices through the resources allocated to each. This experiment complements it by differentiating configurations through their concurrency policy, that is, how many requests each configuration admits for execution at once, as shown by the target and maximum concurrency settings in Table~\ref{tab:benchmark_results}. The three configurations, C1, C2, and C3, produced clearly separated operating points in both throughput and latency. C1 sustained approximately 698--699 tokens/s across all tested loads, while maintaining mean \TTFT close to 100 ms and P99 \TTFT below 140 ms. C2 increased throughput to 883--902 tokens/s, representing a 26--29\% improvement over C1, while maintaining P99 \TTFT below the 500 ms interactive serving threshold at all tested request rates. C3 prioritised capacity, sustaining 2,117--2,130 tokens/s, approximately 2.4$\times$ higher than C2 and 3.0$\times$ higher than C1. Importantly, throughput remained stable once each configuration reached saturation, indicating that the available GPU resources were fully utilised and that additional load was absorbed through queuing rather than causing execution throughput degradation. Furthermore, no request failures were observed in any experiment, demonstrating stable operation across all configurations.

A clear latency-throughput trade-off emerges across the three configurations. C1 consistently delivered the lowest latency, with mean \TTFT remaining between 99 and 101 ms and P99 \TTFT between 130 and 138 ms regardless of load. C2 accepted a moderate increase in latency, with mean \TTFT ranging from 176 to 184 ms and P99 \TTFT between 313 and 472 ms, while still satisfying interactive-service requirements. In contrast, C3 achieved the highest throughput but exhibited significant tail-latency degradation under increasing load. While P99 \TTFT was only 308 ms at 8 RPS, it increased to 1,305 ms at 12 RPS and reached 2,001 ms at 32 RPS, about four times the interactive-service threshold. Notably, \TPOT remained largely constant within each configuration, varying by less than 4\% across the evaluated load range. This indicates that model execution performance remained stable and that the observed latency degradation originated primarily from request admission and queuing effects rather than from reduced decoding efficiency.

Together with the resource-based differentiation in Section~\ref{sec:evaluation:tiers}, these results support the central hypothesis of InferSlice: service-level objectives can be translated into distinct and enforceable operating regimes over a common distributed infrastructure, and concurrency policy provides a second control for doing so. The observed behaviours reveal three practical service classes. C1 provides a latency-first regime suitable for highly interactive applications where responsiveness is the dominant requirement. C2 offers a balanced regime that increases resource utilisation while preserving bounded latency suitable for quality-critical workloads. C3 delivers a throughput-first regime that maximises infrastructure efficiency and is therefore suitable for asynchronous or batch-oriented inference workloads. More broadly, the results demonstrate that AI inference services can be differentiated in a manner analogous to telecom network services, allowing operators to offer multiple \SLO-driven service tiers from the same pool of compute and network resources. This capability is fundamental for future telco AI platforms, where heterogeneous applications must coexist while receiving performance characteristics aligned with their specific operational requirements.

\begin{table*}
\centering
\caption{Inference results across three InferSlice concurrency policies,
         demonstrating distinct latency--throughput operating regimes.
         Policy~C1 enforces a strict concurrency limit suited to
         conversational, latency-sensitive workloads;
         Policy~C2 enforces a conservative queuing policy suited
         to quality-critical workloads; Policy~C3 maximises
         throughput for batch processing. All runs use
         Llama-3.1-8B-Instruct, ShareGPT dataset, Poisson arrivals.
         Zero failed requests across all configurations.}
\label{tab:benchmark_results}
\begin{tabular}{llccccc}
\toprule
\textbf{Policy} & \textbf{RPS} &
\textbf{Mean TTFT (ms)} & \textbf{P99 TTFT (ms)} &
\textbf{Mean TPOT (ms)} & \textbf{Total (tok/s)} &
\textbf{P99 $\leq$ 500~ms} \\
\midrule
\multirow{5}{*}{\shortstack[l]{C1\\{\scriptsize target=1}\\{\scriptsize max=8}\\{\scriptsize concur=8}}}
                   & 4  & 101.07 & 131.33 & 23.07 & 698 & Yes \\
                   & 6  & 99.71  & 131.74 & 23.06 & 699 & Yes \\
                   & 8  & 99.34  & 129.93 & 23.04 & 698 & Yes \\
                   & 10 & 99.79  & 138.18 & 23.08 & 698 & Yes \\
                   & 12 & 99.86  & 130.83 & 23.06 & 698 & Yes \\
\midrule
\multirow{5}{*}{\shortstack[l]{C2\\{\scriptsize target=2}\\{\scriptsize max=16}}}
                   & 4  & 181 & 472 & 35.8 & 900  & Yes \\
                   & 6  & 181 & 451 & 35.9 & 899  & Yes \\
                   & 10 & 184 & 399 & 36.1 & 894  & Yes \\
                   & 12 & 181 & 411 & 36.7 & 883  & Yes \\
                   & 16 & 176 & 313 & 36.0 & 902  & Yes \\
\midrule
\multirow{5}{*}{\shortstack[l]{C3\\{\scriptsize target=4}\\{\scriptsize max=32}}}
                   & 8  & 171 & 308  & 28.2 & 2117 & Yes \\
                   & 12 & 197 & 1305 & 28.4 & 2130 & No  \\
                   & 16 & 210 & 1621 & 28.7 & 2121 & No  \\
                   & 24 & 218 & 1761 & 29.2 & 2122 & No  \\
                   & 32 & 227 & 2001 & 28.7 & 2122 & No  \\
\bottomrule
\end{tabular}
\end{table*}

\subsection{Network Bottleneck Identification}
\subsubsection{Experiment N1 results: End client bandwidth limiting}
\label{app:bw_sweep}

\begin{table*}
\centering
\caption{Client-edge bandwidth sensitivity at client concurrency 10.
The no cap configuration represents the effectively unconstrained
baseline. Bold values highlight the bandwidth transition region and
resulting SLO violations.}
\label{tab:bw_sweep_c10}
\scriptsize
\setlength{\tabcolsep}{3.0pt}
\resizebox{\textwidth}{!}{%
\begin{tabular}{llrrrrrrrrrrr}
\toprule
& &
\multicolumn{2}{c}{\textbf{P99 latency}} &
\textbf{Throughput} &
\textbf{SLO} &
\multicolumn{2}{c}{\textbf{Measured link rate}} &
\multicolumn{2}{c}{\textbf{Mean latency}} &
\multicolumn{2}{c}{\textbf{Median latency}} &
\\
\cmidrule(lr){3-4}
\cmidrule(lr){7-8}
\cmidrule(lr){9-10}
\cmidrule(lr){11-12}
\textbf{Tier} &
\textbf{BW cap} &
\textbf{TTFT} &
\textbf{TPOT} &
\textbf{(tok/s)} &
\textbf{met} &
\textbf{Rx} &
\textbf{Tx} &
\textbf{TTFT} &
\textbf{TPOT} &
\textbf{TTFT} &
\textbf{TPOT} &
\textbf{RPS} \\
&
\textbf{(Gbps)} &
\textbf{(ms)} &
\textbf{(ms)} &
&
&
\textbf{(Mbps)} &
\textbf{(Mbps)} &
\textbf{(ms)} &
\textbf{(ms)} &
\textbf{(ms)} &
\textbf{(ms)} &
\\
\midrule

Premium
& No cap
& 277.06 & 37.99 & 371.97 & Yes
& 0.9100 & 0.1200
& 98.50 & 24.65
& 76.81 & 24.60
& 100.0 \\

Premium
& \textbf{$1\times10^{-3}$}
& 239.16 & 37.32 & 369.14 & Yes
& 0.9200 & 0.1200
& 96.09 & 24.79
& 76.45 & 24.70
& 100.0 \\

Premium
& $5\times10^{-4}$
& \textbf{3276.43} & \textbf{256.86} & 221.52 & \textbf{No}
& 0.4600 & 0.0200
& 1237.59 & 53.71
& 898.96 & 41.38
& 100.0 \\

\midrule

Standard
& No cap
& 182.66 & 26.56 & 377.97 & Yes
& 0.9400 & 0.1200
& 101.76 & 24.11
& 84.28 & 24.65
& 100.0 \\

Standard
& \textbf{$1\times10^{-3}$}
& 405.81 & 26.87 & 379.55 & Yes
& 0.9100 & 0.1200
& 121.50 & 24.25
& 130.72 & 24.65
& 100.0 \\

Standard
& $5\times10^{-4}$
& \textbf{3338.20} & \textbf{276.36} & 216.53 & \textbf{No}
& 0.4600 & 0.0200
& 1402.59 & 56.08
& 932.11 & 41.83
& 100.0 \\

\midrule

Free
& No cap
& 381.11 & 26.39 & 371.59 & Yes
& 0.9000 & 0.1200
& 110.47 & 24.23
& 121.71 & 24.93
& 100.0 \\

Free
& \textbf{$1\times10^{-3}$}
& 377.27 & 26.35 & 377.12 & Yes
& 0.9300 & 0.1200
& 112.89 & 24.18
& 121.54 & 24.91
& 100.0 \\

Free
& $5\times10^{-4}$
& \textbf{1913.70} & \textbf{150.15} & 224.29 & \textbf{No}
& 0.4610 & 0.0289
& 948.83 & 50.85
& 872.36 & 42.71
& 100.0 \\

\bottomrule
\end{tabular}%
}
\end{table*}

\begin{table*}
\centering
\caption{Client-edge bandwidth sensitivity at client concurrency 16.
Premium remains within the TTFT SLO until the available bandwidth is
reduced to 0.5\,Mbps, whereas Standard and Free exhibit TTFT
violations even under effectively unconstrained bandwidth, indicating
queueing or compute-side saturation prior to network saturation.}
\label{tab:bw_sweep_c16}
\scriptsize
\setlength{\tabcolsep}{3.0pt}
\resizebox{\textwidth}{!}{%
\begin{tabular}{llrrrrrrrrrrr}
\toprule
& &
\multicolumn{2}{c}{\textbf{P99 latency}} &
\textbf{Throughput} &
\textbf{SLO} &
\multicolumn{2}{c}{\textbf{Measured link rate}} &
\multicolumn{2}{c}{\textbf{Mean latency}} &
\multicolumn{2}{c}{\textbf{Median latency}} &
\\
\cmidrule(lr){3-4}
\cmidrule(lr){7-8}
\cmidrule(lr){9-10}
\cmidrule(lr){11-12}
\textbf{Tier} &
\textbf{BW cap} &
\textbf{TTFT} &
\textbf{TPOT} &
\textbf{(tok/s)} &
\textbf{met} &
\textbf{Rx} &
\textbf{Tx} &
\textbf{TTFT} &
\textbf{TPOT} &
\textbf{TTFT} &
\textbf{TPOT} &
\textbf{RPS} \\
&
\textbf{(Gbps)} &
\textbf{(ms)} &
\textbf{(ms)} &
&
&
\textbf{(Mbps)} &
\textbf{(Mbps)} &
\textbf{(ms)} &
\textbf{(ms)} &
\textbf{(ms)} &
\textbf{(ms)} &
\\
\midrule

Premium
& No cap
& 418.16 & 39.49 & 543.06 & Yes
& 1.3200 & 0.1700
& 113.24 & 25.91
& 115.29 & 25.39
& 100.0 \\

Premium
& \textbf{$1.5\times10^{-3}$}
& 321.93 & 39.50 & 535.87 & Yes
& 1.3100 & 0.1700
& 109.36 & 26.26
& 115.67 & 25.61
& 100.0 \\

Premium
& $5\times10^{-4}$
& \textbf{4611.48} & \textbf{455.49} & 221.45 & \textbf{No}
& 0.4500 & 0.0200
& 2090.22 & 85.43
& 1432.84 & 66.85
& 100.0 \\

\midrule

Standard
& No cap
& \textbf{508.64} & 33.68 & 565.72 & \textbf{No}
& 1.3800 & 0.1800
& 182.10 & 25.46
& 179.52 & 25.74
& 100.0 \\

Standard
& \textbf{$1.5\times10^{-3}$}
& \textbf{517.36} & 35.39 & 549.58 & \textbf{No}
& 1.3200 & 0.1700
& 173.26 & 25.49
& 179.17 & 25.79
& 100.0 \\

Standard
& $5\times10^{-4}$
& \textbf{3213.88} & \textbf{259.14} & 222.14 & \textbf{No}
& 0.4500 & 0.0300
& 1579.19 & 80.19
& 1441.83 & 68.09
& 100.0 \\

\midrule

Free
& No cap
& \textbf{4547.57} & 27.17 & 437.21 & \textbf{No}
& 1.0400 & 0.1400
& 1860.81 & 24.63
& 1675.30 & 25.20
& 100.0 \\

Free
& \textbf{$1\times10^{-3}$}
& \textbf{4541.20} & 27.58 & 438.43 & \textbf{No}
& 1.0500 & 0.1400
& 1855.40 & 24.67
& 1683.85 & 25.23
& 100.0 \\

Free
& $5\times10^{-4}$
& \textbf{2948.25} & \textbf{248.18} & 224.22 & \textbf{No}
& 0.4600 & 0.0300
& 1502.05 & 81.67
& 1382.30 & 67.99
& 100.0 \\

\bottomrule
\end{tabular}%
}
\end{table*}

\begin{table*}
\centering
\caption{Client-edge bandwidth sensitivity at client concurrency 32.
Premium sustains its unconstrained operating regime at approximately
2.5\,Mbps, while severe bandwidth restriction to 0.5\,Mbps causes
substantial degradation in \TTFT, \TPOT, and token throughput. Standard
and Free are already \TTFT-limited under the unconstrained
baseline, indicating a non-network bottleneck at this load.}
\label{tab:bw_sweep_c32}
\scriptsize
\setlength{\tabcolsep}{3.0pt}
\resizebox{\textwidth}{!}{%
\begin{tabular}{llrrrrrrrrrrr}
\toprule
& &
\multicolumn{2}{c}{\textbf{P99 latency}} &
\textbf{Throughput} &
\textbf{SLO} &
\multicolumn{2}{c}{\textbf{Measured link rate}} &
\multicolumn{2}{c}{\textbf{Mean latency}} &
\multicolumn{2}{c}{\textbf{Median latency}} &
\\
\cmidrule(lr){3-4}
\cmidrule(lr){7-8}
\cmidrule(lr){9-10}
\cmidrule(lr){11-12}
\textbf{Tier} &
\textbf{BW cap} &
\textbf{TTFT} &
\textbf{TPOT} &
\textbf{(tok/s)} &
\textbf{met} &
\textbf{Rx} &
\textbf{Tx} &
\textbf{TTFT} &
\textbf{TPOT} &
\textbf{TTFT} &
\textbf{TPOT} &
\textbf{RPS} \\
&
\textbf{(Gbps)} &
\textbf{(ms)} &
\textbf{(ms)} &
&
&
\textbf{(Mbps)} &
\textbf{(Mbps)} &
\textbf{(ms)} &
\textbf{(ms)} &
\textbf{(ms)} &
\textbf{(ms)} &
\\
\midrule

Premium
& No cap
& 379.14 & 44.51 & 1001.40 & Yes
& 2.4500 & 0.3600
& 131.10 & 30.25
& 125.55 & 29.88
& 100.0 \\

Premium
& \textbf{$2.5\times10^{-3}$}
& 416.14 & 47.14 & 933.89 & Yes
& 2.3500 & 0.3400
& 132.70 & 30.67
& 124.12 & 29.91
& 100.0 \\

Premium
& $5\times10^{-4}$
& \textbf{12772.72} & \textbf{1187.84} & 231.08 & \textbf{No}
& 0.4700 & 0.0300
& 4564.35 & 178.11
& 4393.32 & 124.69
& 100.0 \\

\midrule

Standard
& No cap
& \textbf{7876.05} & 40.72 & 617.85 & \textbf{No}
& 1.4400 & 0.2000
& 4308.16 & 26.56
& 4211.11 & 27.21
& 100.0 \\

Standard
& \textbf{$1.5\times10^{-3}$}
& \textbf{7480.05} & 136.18 & 604.78 & \textbf{No}
& 1.3400 & 0.0900
& 3400.20 & 40.59
& 3277.30 & 32.93
& 100.0 \\

Standard
& $5\times10^{-4}$
& \textbf{7291.51} & \textbf{537.05} & 223.35 & \textbf{No}
& 0.4500 & 0.0300
& 3171.58 & 157.86
& 2789.21 & 135.70
& 100.0 \\

\midrule

Free
& No cap
& \textbf{12716.79} & 31.98 & 446.12 & \textbf{No}
& 1.0800 & 0.1400
& 8110.86 & 23.64
& 8070.19 & 24.17
& 100.0 \\

Free
& \textbf{$1\times10^{-3}$}
& \textbf{12540.49} & 26.91 & 446.78 & \textbf{No}
& 1.1000 & 0.1400
& 8107.47 & 23.55
& 8172.76 & 24.26
& 100.0 \\

Free
& $5\times10^{-4}$
& \textbf{14120.28} & \textbf{553.79} & 231.38 & \textbf{No}
& 0.4500 & 0.0300
& 5031.64 & 117.36
& 2843.13 & 125.58
& 100.0 \\

\bottomrule
\end{tabular}%
}
\end{table*}

At moderate load (concurrency 10), all three service tiers remain largely insensitive to bandwidth while the available link capacity exceeds approximately 1~Mbps~(Table~\ref{tab:bw_sweep_c10}).
Reducing the cap from 100,Gbps to 1~Mbps produces little change in throughput or \TPOT:~measured receive rates remain around 0.9--0.94~Mbps and throughput around 370--380~tok/s, while P99 \TTFT remains below the 500~ms \SLO.
A clear network bottleneck emerges when the cap is reduced to 0.5~Mbps.
At this point, the measured receive rate saturates at approximately 0.46~Mbps across all tiers, token throughput falls to 216--224~tok/s, and P99 \TTFT increases from hundreds of milliseconds to 1.9--3.3~s. 
\TPOT simultaneously rises from approximately 26--38~ms to 150--276~ms.
This sharp transition indicates a client-edge bandwidth floor between 0.5 and 1~Mbps for this workload: above the floor, inference and scheduling dominate latency; below it, network throughput becomes the binding resource and affects both first-token delivery and subsequent token generation.

As concurrency increases, the bottleneck shifts, as shown in Tables~\ref{tab:bw_sweep_c16} and~\ref{tab:bw_sweep_c32}.
For Premium, the unconstrained traffic demand rises from approximately 0.91~Mbps at concurrency 10 to 1.32~Mbps at 16 and 2.45~Mbps at 32, and correspondingly the tier remains close to its unconstrained operating point under caps of 1, 1.5, and 2.5~Mbps, respectively.
In every case, constraining the link to 0.5~Mbps forces the measured receive rate to approximately 0.45--0.47~Mbps and collapses throughput to roughly 221--231~tok/s; at concurrency 32, Premium P99 \TTFT reaches 12.8~s and P99 \TPOT 1.19~s~(Table~\ref{tab:bw_sweep_c32}).
Standard and Free, however, already violate the \TTFT \SLO at higher concurrency even with an effectively unconstrained 100~Gbps link~(Tables~\ref{tab:bw_sweep_c16} and~\ref{tab:bw_sweep_c32}), while \TPOT remains comparatively stable.
Their high \TTFT at these operating points therefore cannot be attributed primarily to network bandwidth and instead reflects queueing or compute-side saturation.
Bandwidth restriction compounds this degradation once the 0.5~Mbps floor is reached.
These results distinguish the network bottleneck from the previously observed queue boundary:~queue saturation predominantly increases \TTFT, whereas severe bandwidth saturation simultaneously reduces measured network throughput and token throughput and substantially increases both \TTFT and \TPOT.

\newpage

\section{List of Abbreviations}
\label{sec:appendix:abbreviations}
\printacronyms[heading=none]

\end{document}